\documentclass[preprint,prd,tightenlines,nofootinbib,superscriptaddress, 11pt]{article}
\usepackage{graphicx,amsmath,amssymb,epsf, fdag, bm}

\newcommand{\tr}{{\rm tr}}

\usepackage[utf8x]{inputenc}
\usepackage[english]{babel}
\usepackage{cite}
\usepackage{color}
\usepackage{graphicx}
\definecolor{red}{rgb}{1,0,0}
\def\lesssim{\ \hbox{\raise 2pt \hbox{$<$} \kern -13pt
                     \lower 3pt \hbox{$\sim$}}\ }
\def\greatersim{\ \hbox{\raise 2pt \hbox{$>$} \kern -13pt
                     \lower 3pt \hbox{$\sim$}}\ }
\input epsf.tex
\def\desepsf(#1 width #2){\epsfxsize=#2 \epsfbox{#1}}

\begin{document}

\title{\bf \Large  Forward jet production \& quantum corrections to the gluon Regge trajectory from Lipatov's high energy effective action}
\author{G.\ Chachamis${}^{1}$, M.\ Hentschinski${}^{2}$, J.~D.\ Madrigal Mart\'inez${}^{3}$, A.~Sabio Vera${}^{3}$ } 

\maketitle

\begin{center}
${}^{1}$ Instituto de F{\' \i}sica Corpuscular, Consejo Superior de Investigaciones Cient{\' \i}ficas-Universitat de Val{\`e}ncia, Parc Cient{\' \i}fic, E-46980 Paterna, Valencia, Spain. \\ 
  ${}^{2}$ Physics Department, Brookhaven National Laboratory,   Upton, NY, 11973, USA. \\
${}^3$ Instituto de F{\' i}sica Te{\' o}rica  
UAM/CSIC \& Universidad Aut{\' o}noma de Madrid, \\ E-28049 Madrid, Spain.
\end{center}

\begin{abstract}

We review Lipatov's high energy effective action and show that it is a useful computational tool to calculate scattering amplitudes in (quasi)-multi-Regge kinematics. We explain in some detail our recent work where a novel regularization and subtraction procedure has been proposed that allows to extend the use of this effective action beyond tree level. Two 
examples are calculated at next-to-leading order: forward jet vertices and the gluon Regge trajectory. 
\end{abstract}

\section{Introduction} 

A useful description of the high energy limit of perturbative QCD can be given 
in terms of high energy factorization and the BFKL evolution equation. 
In this framework large logarithms in the center of mass
energy $\sqrt{s}$ at leading (LL)~\cite{Fadin:1975cb}
and next-to-leading logarithmic  (NLL) accuracy \cite{BFKLNLO} are resummed. 
Applications of this framework to LHC and HERA phenomenology are numerous, for recent results see \cite{de}.
The building block in this formalism is the realization that QCD scattering
amplitudes in the high energy limit are dominated by the exchange of an effective $t$-channel
degree of freedom, the so-called reggeized gluon, which couples to the external
scattering particles through process dependent effective vertices.  The calculation of higher order corrections 
to both the effective couplings and the reggeized gluon propagators is a difficult task. This
is especially true when attempting to go beyond LL accuracy, with the aim to 
obtain accurate phenomenological predictions. 

A powerful tool to easy these calculations is provided by Lipatov's
high energy effective action~\cite{LevSeff}, which corresponds to the
usual QCD action with the addition of an induced term containing the
new effective degrees of freedom relevant in this limit.  This induced
piece is written in terms of gauge-invariant currents which generate a
non-trivial coupling of the gluon to the reggeized gluon fields. Tree
level amplitudes at high energies have been obtained making use of
this action in Ref.~\cite{Antonov:2004hh}.  Loop corrections are
technically more involved since a new type of longitudinal
divergences, not present in conventional QCD amplitudes, appear.  The
treatment of these divergences has been addressed for LL transition
kernels in Ref.~\cite{Martin} while the coupling for one to two
reggeized gluons with the associated production of an on-shell gluon
has been studied in~\cite{Braun}.
Recently, this high energy effective action has been used for the
first time for the calculation of NLL corrections to the forward
quark-initiated jet vertex~\cite{Hentschinski:2011tz} and the quark
contribution to the two-loop gluon trajectory~\cite{Chachamis:2012gh},
finding exact agreement with previous results in the literature. In
the following we review the key steps to perform these calculations.

\section{High energy factorization and the effective action}
\label{sec:action}

Following Lipatov's work~\cite{LevSeff} in this section we motivate the construction of
gauge invariant high energy factorization and the effective action from an explicit study of QCD tree-level amplitudes. 

To fix our notation let us use the partonic scattering process 
$p_A + p_B \to p_1 + p_2 + \ldots$, with light-like initial momenta $p_A^2 = p_B^2 = 0$ and squared
center of mass energy $s = 2 p_A \cdot p_B$. Dimensionless
light-cone four-vectors $n^+$ and $n^-$ are then defined through the following 
rescaling $n^+ = 2 p_B/ \sqrt{s}$ and $n^- = 2 p_A/\sqrt{s} $.  The general 
Sudakov decomposition of a four vector $k^\mu$ can then be written in the form 
\begin{align}
  \label{eq:lc_comp}
   k^\mu &= \frac{k^+}{2} (n^-)^\mu +  \frac{k^-}{2} (n^+)^\mu + {\bm k}^\mu,
&
 k^\pm & = n^\pm\cdot k, 
&
 n^\pm\cdot {\bm k} = 0.
\end{align}
We start with the simpler example of a scalar $\phi^3$-theory\footnote{For   the formulation of  this model and a comprehensive  analysis of its high energy behavior, we refer the interested reader to the literature, {\it e.g.}  \cite{Gribov:2003nw}}. The tree-level diagrams for a $2 \to 2$
scattering process are
 \begin{align}
\label{eq:phi3}
 \parbox{2.7cm}{ \includegraphics[width=2.7cm]{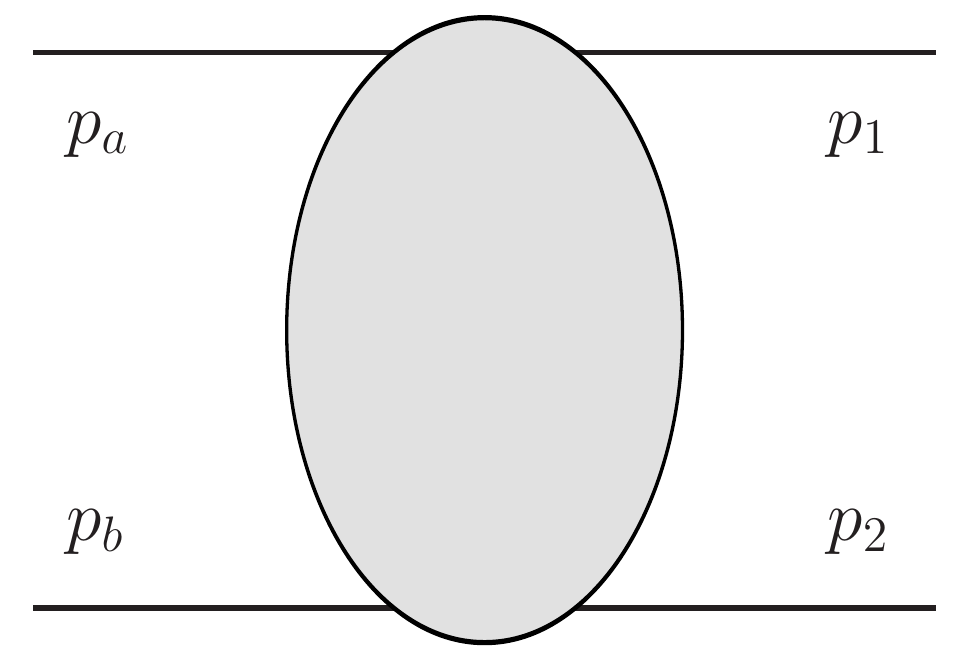}} & = 
\parbox{2cm}{\includegraphics[width = 2.5cm]{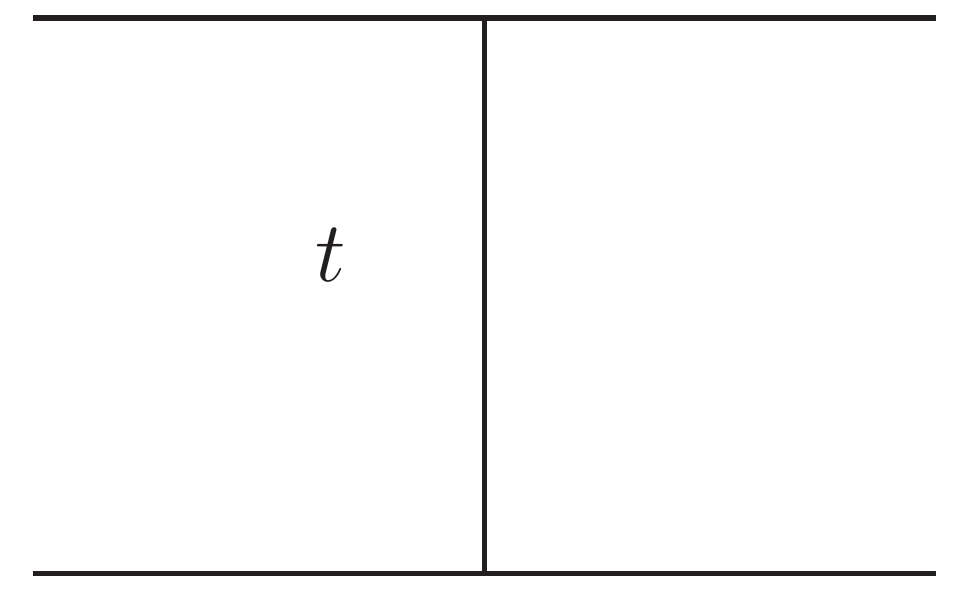}} \quad  + \quad
 \parbox{2cm}{\includegraphics[width = 2.5cm]{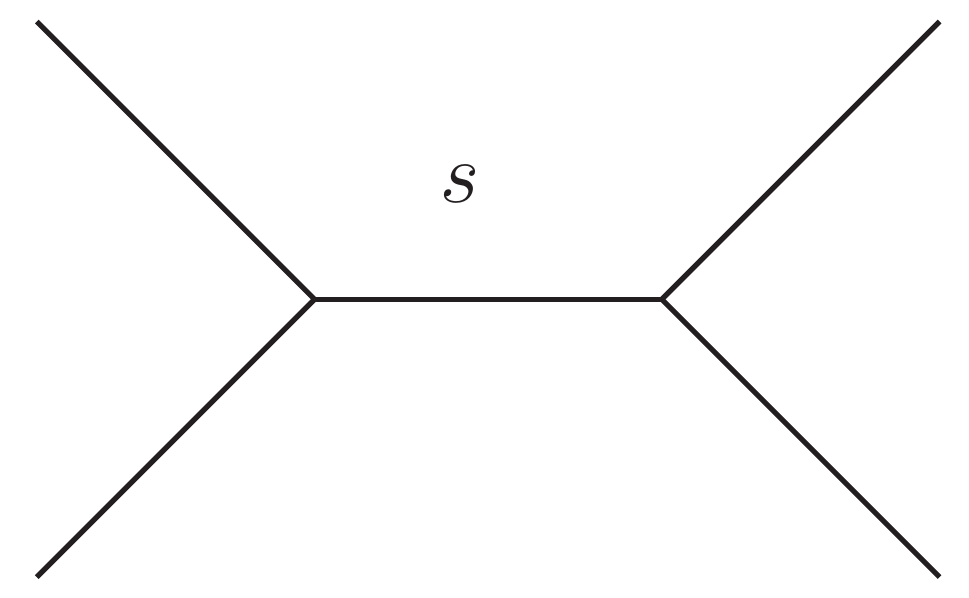}}  \quad + \quad 
  \parbox{2cm}{\includegraphics[width = 2.5cm]{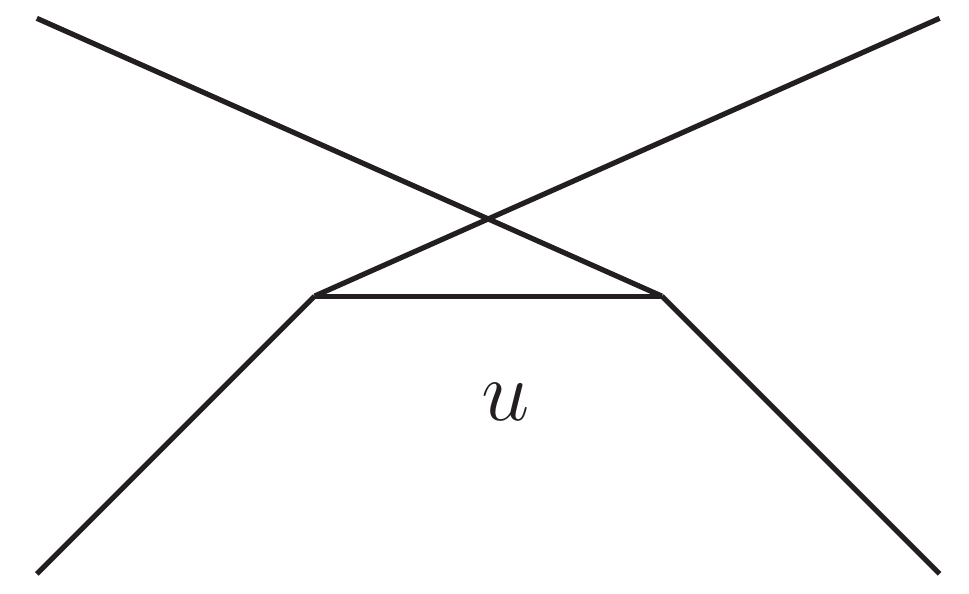}} ~~~~~.
\end{align}
In this case the high energy limit $s \sim - u \gg -t$ with $t = q^2$, $q = p_b - p_2$ and $u =
(p_a - p_2)^2$ is driven by the first diagram of the sum in
Eq.~\eqref{eq:phi3}, while the $s$- and $u$-channel diagrams are power-suppressed with energy. 
In these kinematics  the `plus' (`minus')  momenta  of the upper (lower) particles in the diagram are conserved, $p_a^+ \simeq p_1^+$ and $p_b^- \simeq p_2^-$, which implies that the $t$-channel momentum $q$ takes for the `upper' particles with momenta $p_a$ and $p_1$  the effective form
\begin{align}
  \label{eq:q_regionA}
  q &= \frac{ q^-}{2} n^+ + {\bm q} 
&& \text{(region a-1)},
\end{align}
while for the lower particles with momenta $p_b$ and $p_2$ we have
\begin{align}
  \label{eq:q_regionB}
  q &= \frac{ q^+}{2} n^- + {\bm q} 
 && \text{(region b-2)}.
\end{align}

In QCD a similar process is given by elastic gluon-quark scattering ($g(p_a)q(p_b) \to g(p_1) g(p_2)$) 
with the following diagrams
\begin{align}
  \label{eq:gqqcd}
  \parbox{3cm}{\includegraphics[width = 3cm]{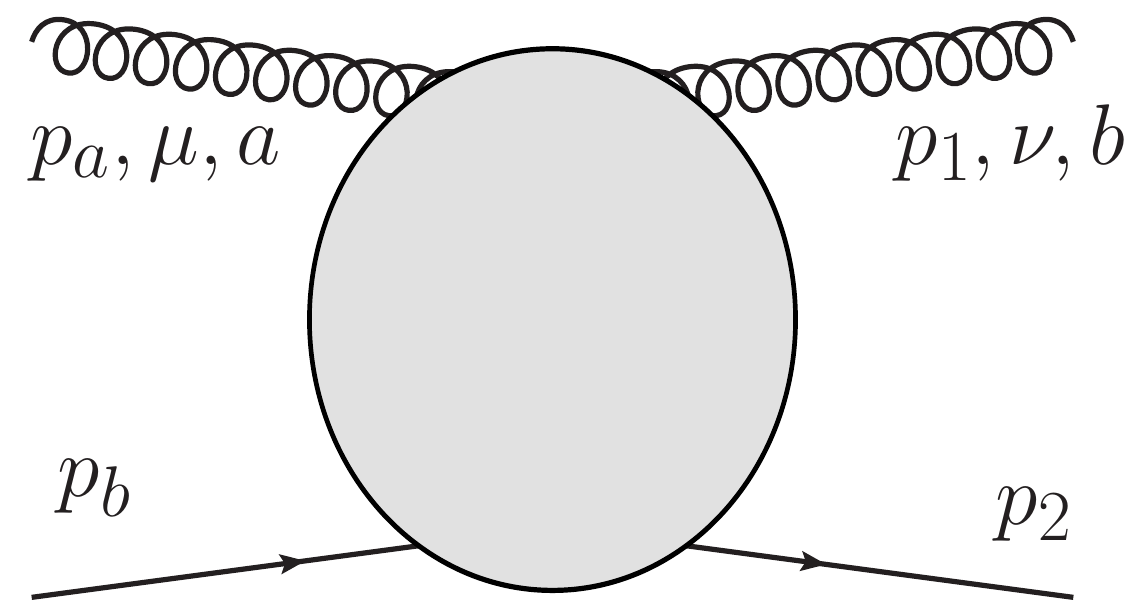}} & = 
 \parbox{3cm}{\includegraphics[width = 3cm]{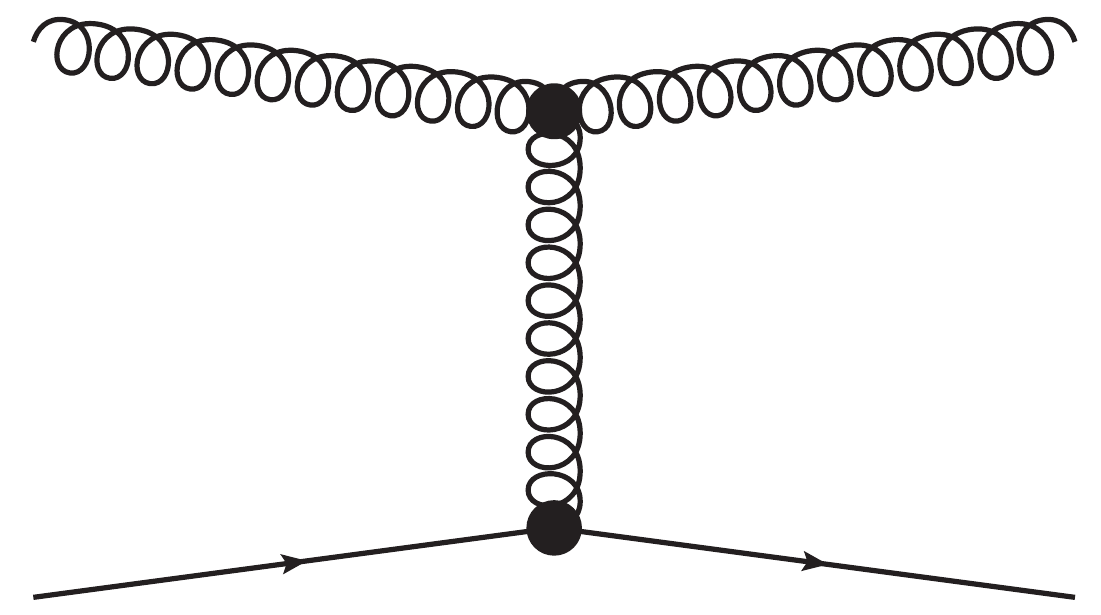}} + 
\parbox{3cm}{\includegraphics[width = 3cm]{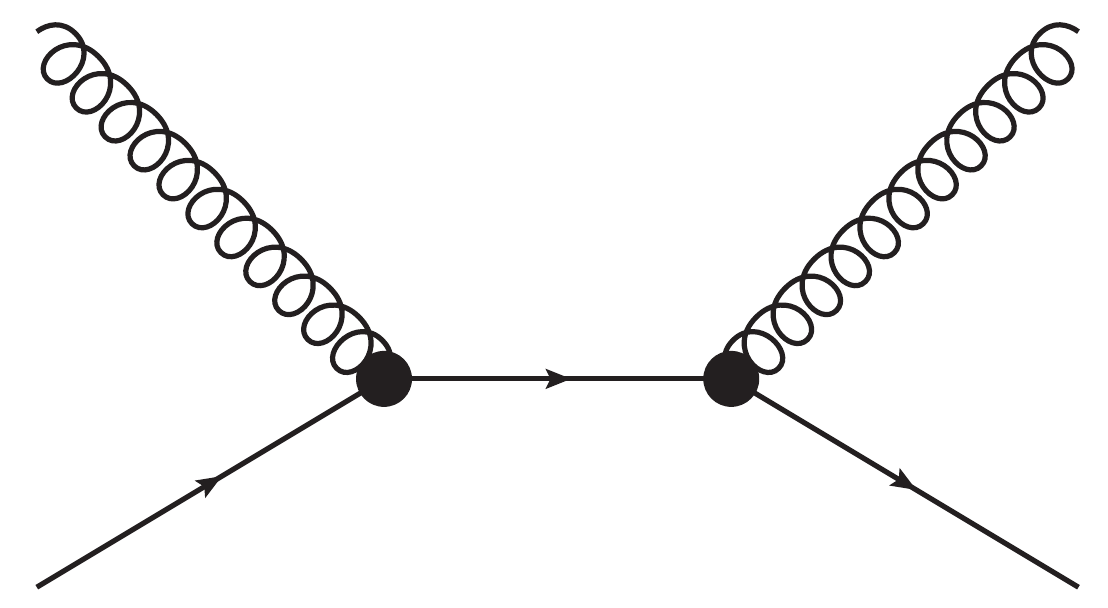}} + 
\parbox{3cm}{\includegraphics[width = 3cm]{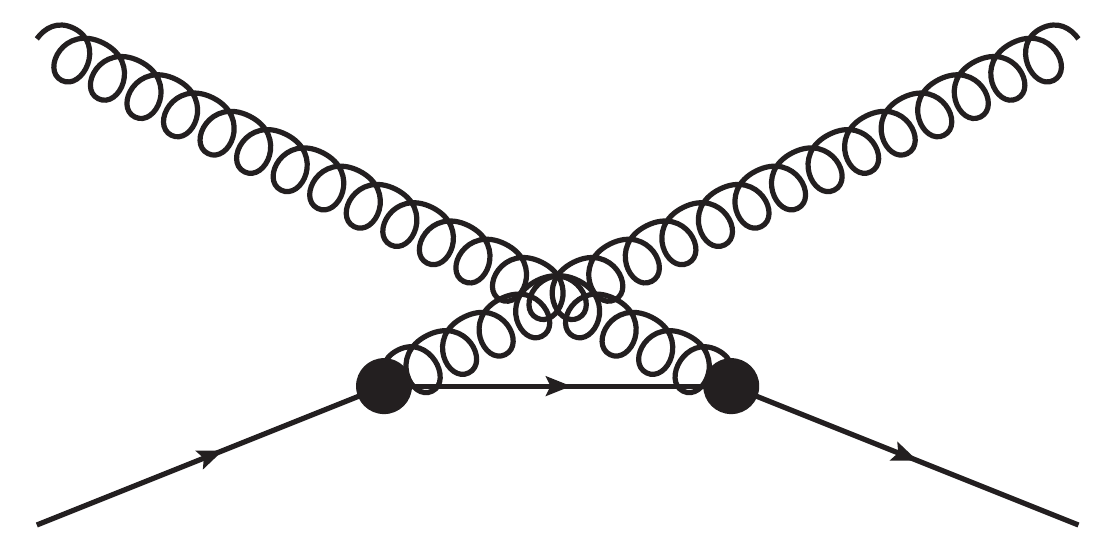}}.
\end{align}
Kinematics carries over from the $\phi^3$-theory,  with the additional feature that for covariant gauges certain longitudinal  components of the gauge fields are enhanced in comparison to their transverse counterparts\footnote{A gauge invariant formulation of this statement is  possible at the level of the gluon field strength tensor (see {\it e.g.} \cite{Weigert:2005us,Balitsky:1995ub} and references therein).}. In particular, in covariant Feynman gauge, the polarization tensor of the $t$-channel gluon is to be replaced by its enhanced longitudinal part $g_{\mu\nu} \to n^+_\mu n^-_\nu/2$ while the $s$-channel gluons emitted directly from the quark line have, up to power suppressed corrections, polarizations proportional to $ n^+_\mu$. 

In contrast to  kinematical effects, for which QCD and the scalar theory
agree, the gauge theory nature of QCD no longer allows to drop the contribution from the $s$- and $u$-channel diagrams.  While certain choices such as the $V \cdot n^+ = 0$
light-cone gauge allow to cast the entire relevant contributions into
the $t$-channel diagram, gauge invariance requires to take into
account the full set of diagrams in Eq.~\eqref{eq:gqqcd}. Na\"ive high
energy factorization at the level of QCD Feynman diagrams therefore no longer occurs.

Within the effective action proposed in \cite{LevSeff} this
problem is solved by introducing an additional effective scalar
degree-of-freedom which describes the interaction between particles
with significantly different rapidity $y = 1/2 \ln (k^+/k^-)$. Due to
its scalar nature, this degree-of-freedom (which is identified with
the reggeized gluon) is invariant under gauge transformations and sets
the basis for gauge invariant high energy factorization. Building on
the $\phi^3$-theory result and on the observation that 
the $V \cdot n^+ = 0$ gauge still allows to describe the entire
process in terms of the $t$-channel contributions only, the effective
action formalism takes as a starting point the QCD $t$-channel topologies. 
As a second step these contributions are dressed, making use of new `induced vertices' ,  
with precisely those contributions of the QCD $s$- and $u$-channel diagrams which are
needed to restore gauge invariance. 

\begin{figure}[htb]
  \centering
   \parbox{.7cm}{\includegraphics[height = 1.8cm]{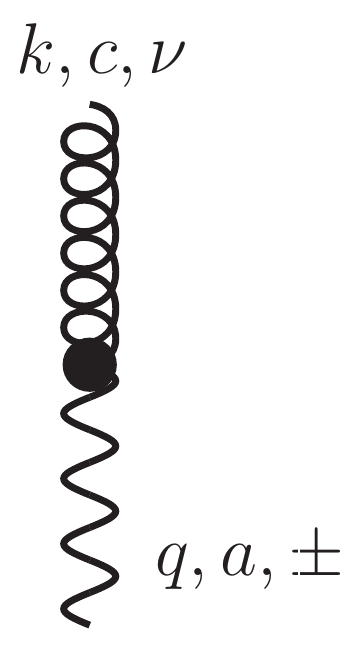}} $=  \displaystyle 
   \begin{array}[h]{ll}
    \\  \\ - i{\bm q}^2 \delta^{a c} (n^\pm)^\nu,  \\ \\  \qquad   k^\pm = 0.
   \end{array}  $  
 \parbox{1.2cm}{ \includegraphics[height = 1.8cm]{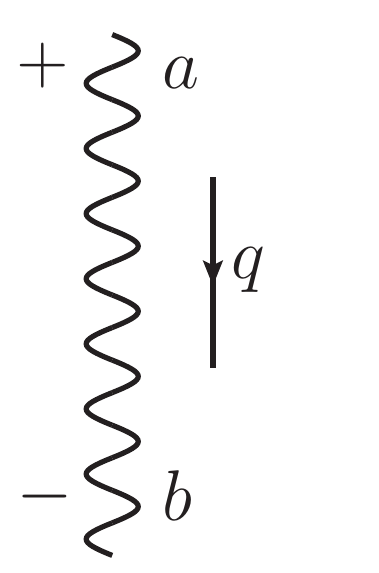}}  $=  \displaystyle    \begin{array}[h]{ll}
    \delta^{ab} \frac{ i/2}{{\bm q}^2} \end{array}$ 
 \parbox{1.7cm}{\includegraphics[height = 1.8cm]{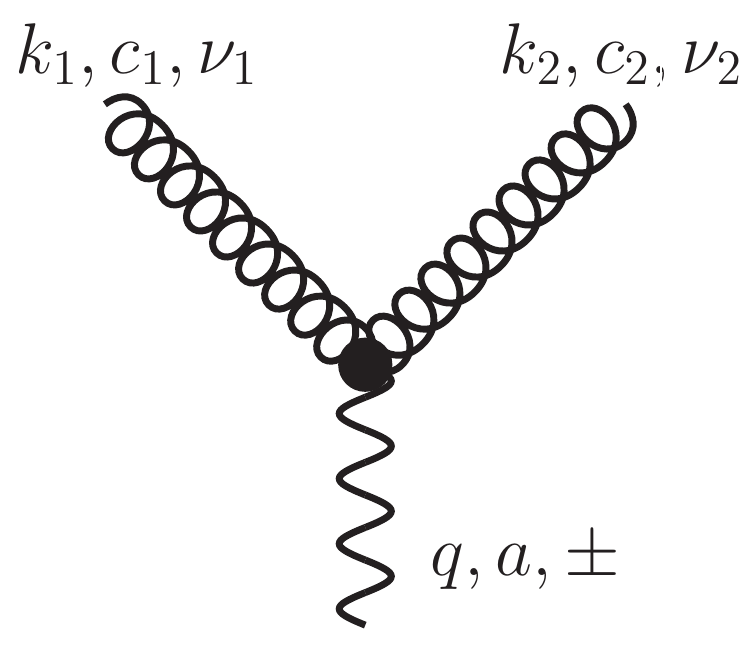}} $ \displaystyle  =  \begin{array}[h]{ll}  \\ \\ g f^{c_1 c_2 a} \frac{{\bm q}^2}{k_1^\pm}   (n^\pm)^{\nu_1} (n^\pm)^{\nu_2},  \\ \\ \quad  k_1^\pm  + k_2^\pm  = 0
 \end{array}$
 \\
\parbox{4cm}{\center (a)} \parbox{4cm}{\center (b)} \parbox{4cm}{\center (c)}
  \caption{\small Transition vertex (a), reggeized gluon propagator (b) and induced vertex (c).}
  \label{fig:feynrules0p2}
\end{figure}

The elastic gluon-quark scattering amplitude in the
high energy limit is then described by the following set of diagrams
\begin{align}
  \label{eq:gqeff}
  \parbox{3cm}{\includegraphics[width = 3cm]{gq_gen.pdf}} & = 
 \parbox{3cm}{\includegraphics[width = 3cm]{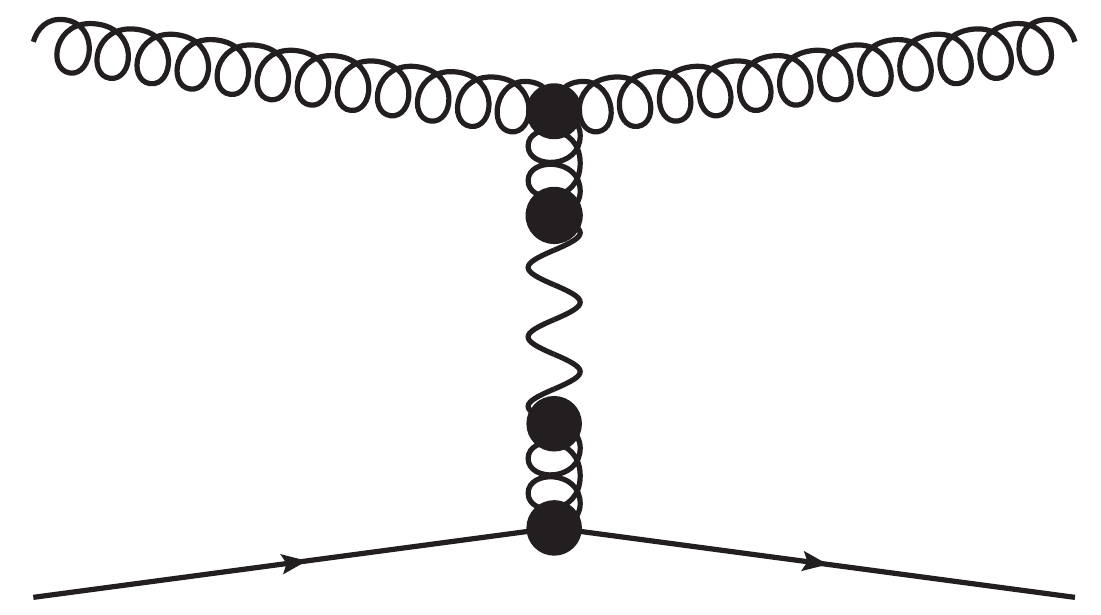}} + 
\parbox{3cm}{\includegraphics[width = 3cm]{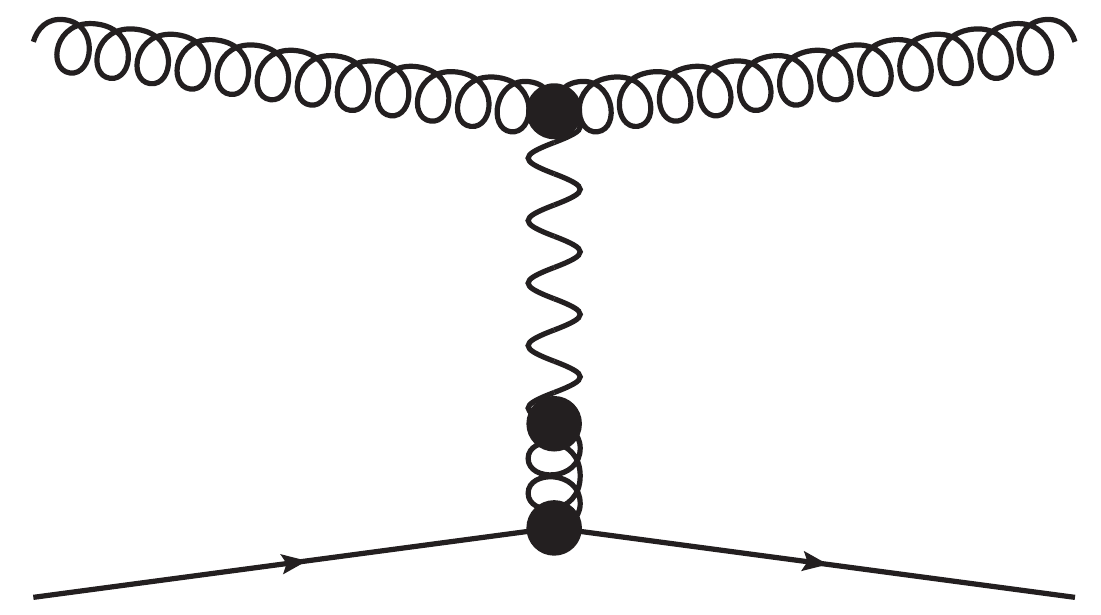}} + \mathcal{O}\left (\frac{|t|}{s} \right).
\end{align}
The Feynman rules needed for the construction of these  diagrams are shown  in Fig.~\ref{fig:feynrules0p2}.

The first element, Fig.~\ref{fig:feynrules0p2}.a, yields  the projection of the QCD gluon on the 
high energy kinematics and polarization. 
Fig.~\ref{fig:feynrules0p2}.b describes the  propagator of the effective
$t$-channel exchange; it agrees with the high energy limit of the $t$-channel gluon propagator in covariant gauge, with polarization vectors $n^+_\mu n^-_\nu$ being absorbed into the adjacent vertices.   Fig.~\ref{fig:feynrules0p2}.c  corresponds to the 
$\mathcal{O}(g)$ \emph{induced vertex}, which makes the two gluon-reggeized gluon amplitude to be gauge invariant. With the latter defined as
\begin{align}
  \label{eq:2gr}
  i\mathcal{M}^{\mu\nu, abc}_{2gr^*} (p_a, p_1, q) & = \parbox{2cm}{\center \includegraphics[height = 1.8cm]{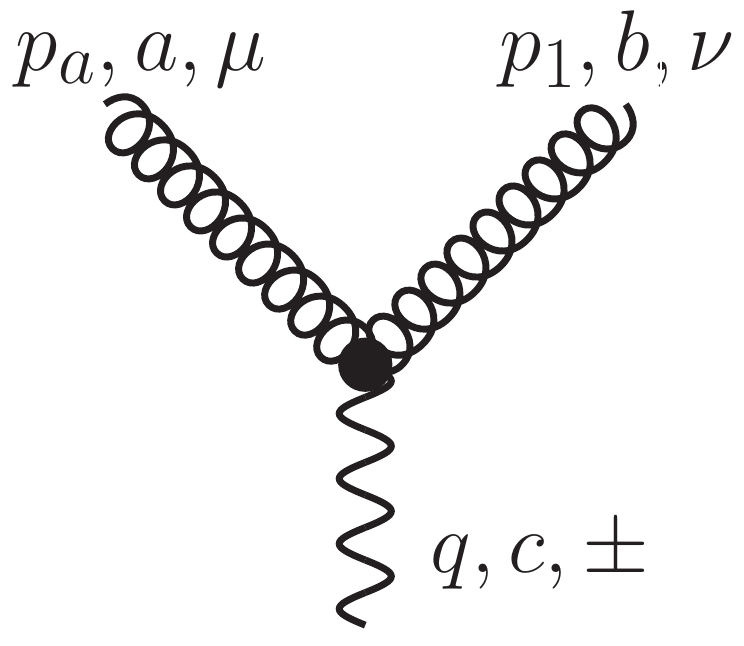}} +
 \parbox{2cm}{\center\includegraphics[height = 1.8cm]{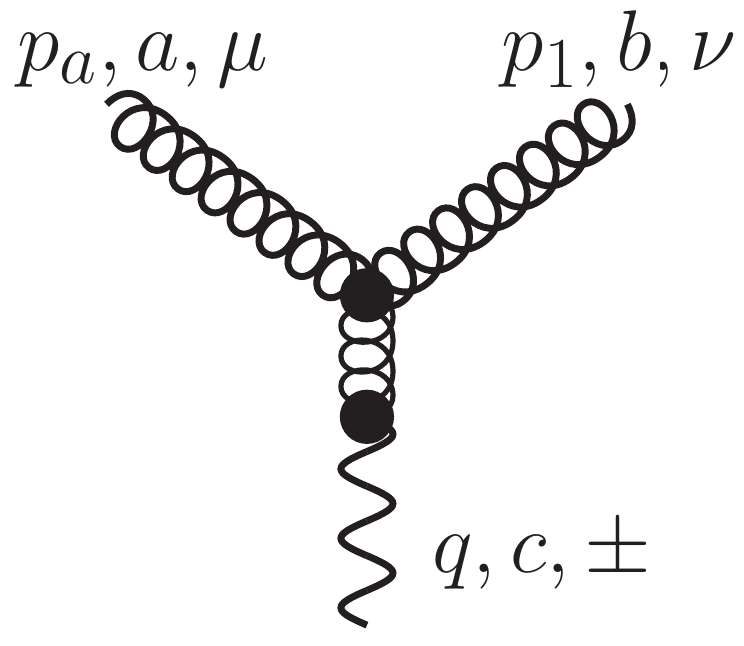}}
\end{align}
one explicitly finds that the corresponding Slavnov-Taylor identities are fulfilled
\begin{align}
  \label{eq:SlavTayl}
  {p_a}_\mu \mathcal{M}^{\mu\nu}_{2gr^*} (p_a, p_1, q) \epsilon^{(\lambda)}_\nu (p_1) & = 0 =  {p_a}_\mu \mathcal{M}^{\mu\nu}_{2gr^*} (p_a, p_1, q) {p_1}_\nu,  
\notag \\
 \epsilon^{(\lambda)}_\nu (p_a) \mathcal{M}^{\mu\nu}_{2gr^*} (p_a, p_1, q)  {p_1}_\nu  & = 0. 
\end{align}

\begin{figure}[htb]
  \centering
  \parbox{.47 \textwidth}{\center \includegraphics[width=5cm]{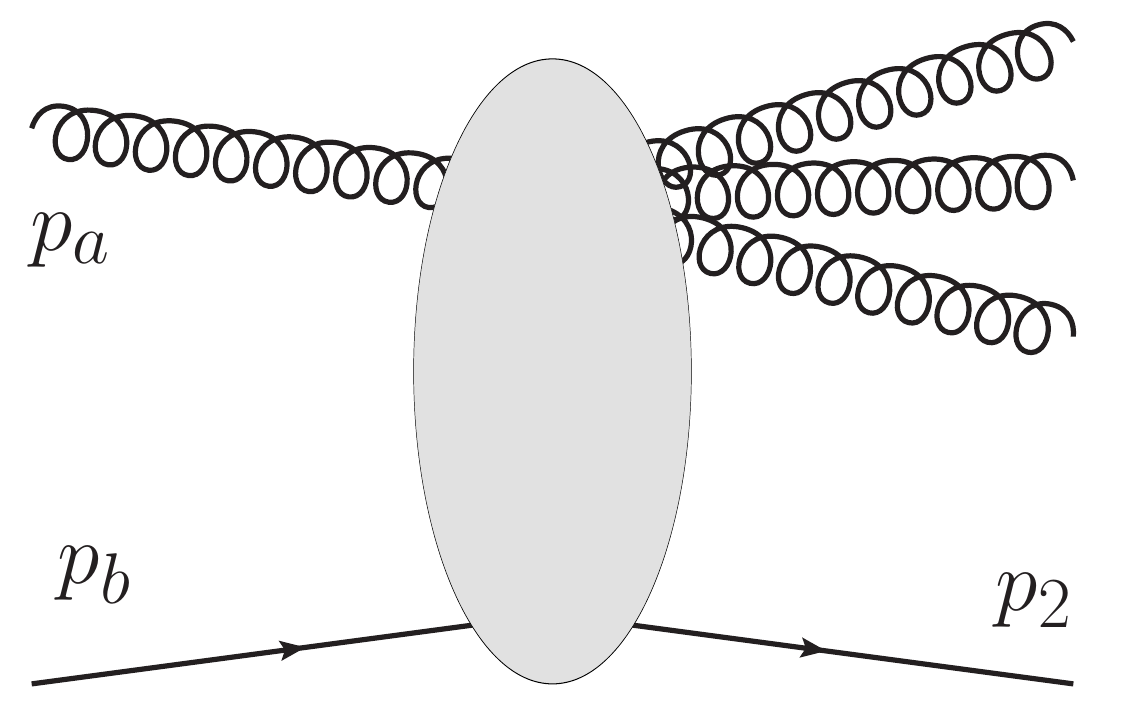}}
 \parbox{.47 \textwidth}{\center \includegraphics[width=5cm]{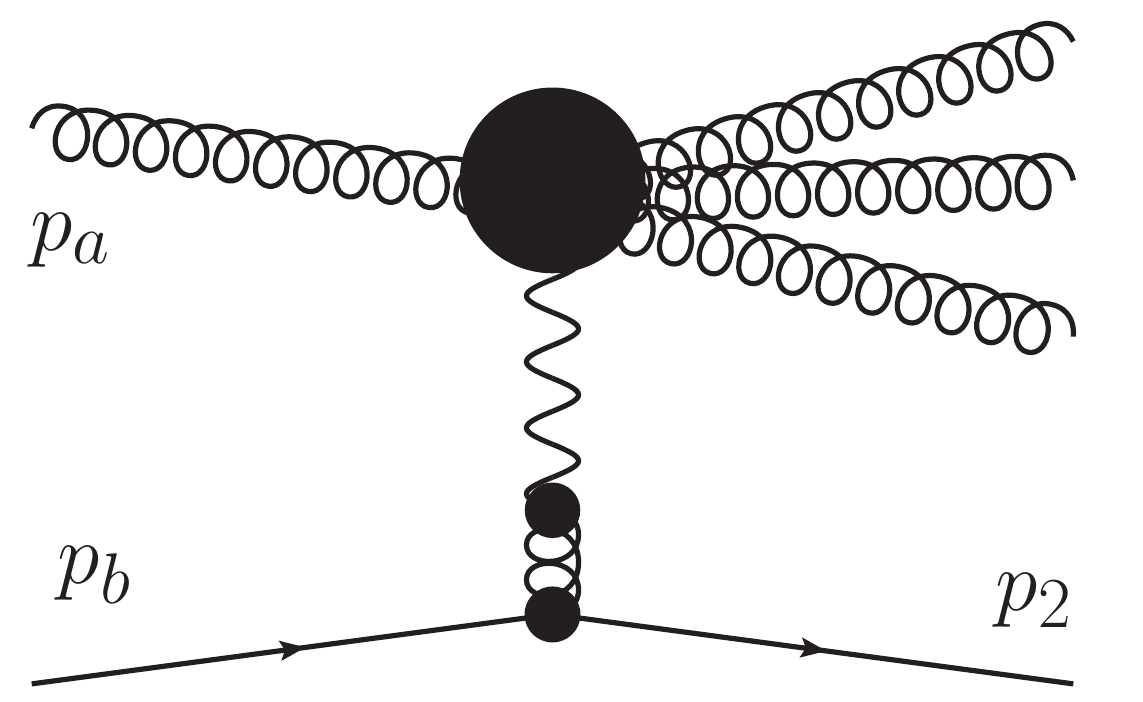}}

 \parbox{.47 \textwidth}{\center (a) }\parbox{.47 \textwidth}{\center (b) }
 \caption{\small a) Quasi-elastic scattering of a gluon on a quark
   with final state gluons produced in the fragmentation region of the
   initial gluon, well-separated separated in rapidity from the 
   target quark. b) The effective action formalism factorizes this
   process with the fragmentation process described by a gauge
   invariant $n$ gluon-reggeized gluon amplitude.  }
  \label{fig:ggg}
\end{figure}

This structure can be extended to the case where $n$-gluons are produced in the fragmentation region of the initial gluon and the whole cluster is well-separated in rapidity from 
the final state quark, see Fig.~\ref{fig:ggg}. 
To achieve gauge invariance for the (tree-level) $n$-gluon--reggeized
gluon amplitude, it is needed to supplement the set in 
Fig.~\ref{fig:feynrules0p2} for the $i$th additional external gluon
with the corresponding $\mathcal{O}(g^{i+1})$ induced vertex. As a
result one obtains a recursive relation for the $\mathcal{O}(g^n)$
induced vertex; the latter can then be expressed as the
$\mathcal{O}(g^n)$ term of a path order exponential. This fact was 
first observed in~\cite{LevSeff, Antonov:2004hh} and has been 
recently rederived in~\cite{vanHameren:2012uj}. 

Lipatov's effective action generates the above-mentioned set of diagrammatic
rules by adding to the QCD action, $S_{\text{QCD}}$, an additional
induced term, $ S_{\text{ind.}}$, which describes the coupling of the
gluonic field $v_\mu = -it^a v_\mu^a(x)$ to the reggeized gluon field
$A_\pm(x) = - i t^a A_\pm^a (x)$.  We therefore have
\begin{align}
  \label{eq:effac}
S_{\text{eff}}& = S_{\text{QCD}} +
S_{\text{ind.}}  ,  
\end{align}
where we include the gauge-fixing and ghost terms
\begin{align}
  \label{eq:action_QCD}
{S}_{\text{QCD}} = 
 \int d^4 x  \big[\mathcal{L}_{\text{QCD}}& (v_\mu, \psi, \bar{\psi}) + \mathcal{L}_{\text{fix}}(v_\mu) + \mathcal{L}_{\text{ghost}}(v_\mu, \phi, \phi^\dagger)\big].
\end{align}
The reggeized gluon field obeys the kinematic constraint
\begin{align}
  \label{eq:kinematic}
  \partial_+ A_- & = 0 = \partial_+ A_+,
\end{align}
in order to fulfill the kinematic conditions in Eqs.~\eqref{eq:q_regionA}
and \eqref{eq:q_regionB}.  Although the reggeized gluon field is
charged under the QCD gauge group SU$(N_c)$, it is invariant under
local gauge transformations, $\delta A_\pm = 0$.  Its kinetic term and
the gauge invariant coupling to the QCD gluon field are included in
the induced term
\begin{align}
\label{eq:1efflagrangian}
  S_{\text{ind.}} = \int \text{d}^4 x
\tr\left[\left(W_-[v] - A_- \right)\partial^2_\perp A_+\right]
+\tr\left[\left(W_+[v] - A_+ \right)\partial^2_\perp A_-\right].
\end{align}
The definition of the  non-local functionals $W_\pm[v]$ reads
\begin{align}
  \label{eq:funct_expand}
  W_\pm[v] =&
- \frac{1}{g} \partial_\pm  U[v_\pm]
= v_\pm - g  v_\pm\frac{1}{\partial_\pm} v_\pm + g^2 v_\pm
\frac{1}{\partial_\pm} v_\pm\frac{1}{\partial_\pm} v_\pm - \ldots
\end{align}
with
\begin{align}
  \label{eq:Wilsonlines}
  U[v_\pm] & =  \mathrm{P} \exp \left(- \frac{g}{2} \int^{x^\pm}_{- \infty} d z^\pm v_\pm(z^\pm, x_\perp) 
\right)
&& \text{and} & x_\perp& = (x^\mp, {\bm x}),
\end{align}
being a path ordered exponential with an integration contour located along the two light-cone
directions and with the boundary condition $\lim_{x^\pm \to \infty} U[v_\pm] = 1$.

\begin{figure}[htb]
  \centering
  \parbox{.47 \textwidth}{\center \includegraphics[width = 3.6cm]{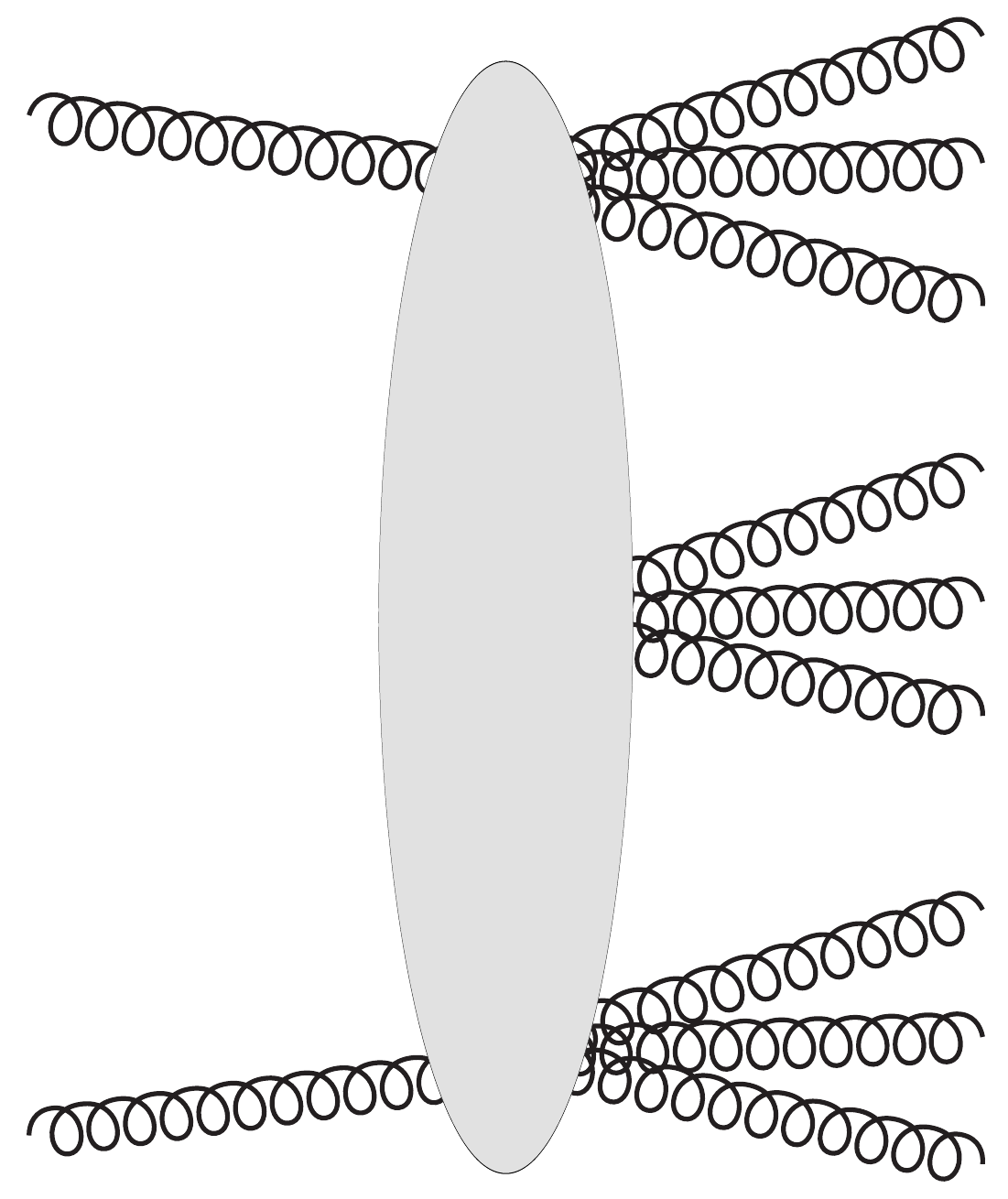}}
 \parbox{.47 \textwidth}{\center \includegraphics[width = 3.6cm]{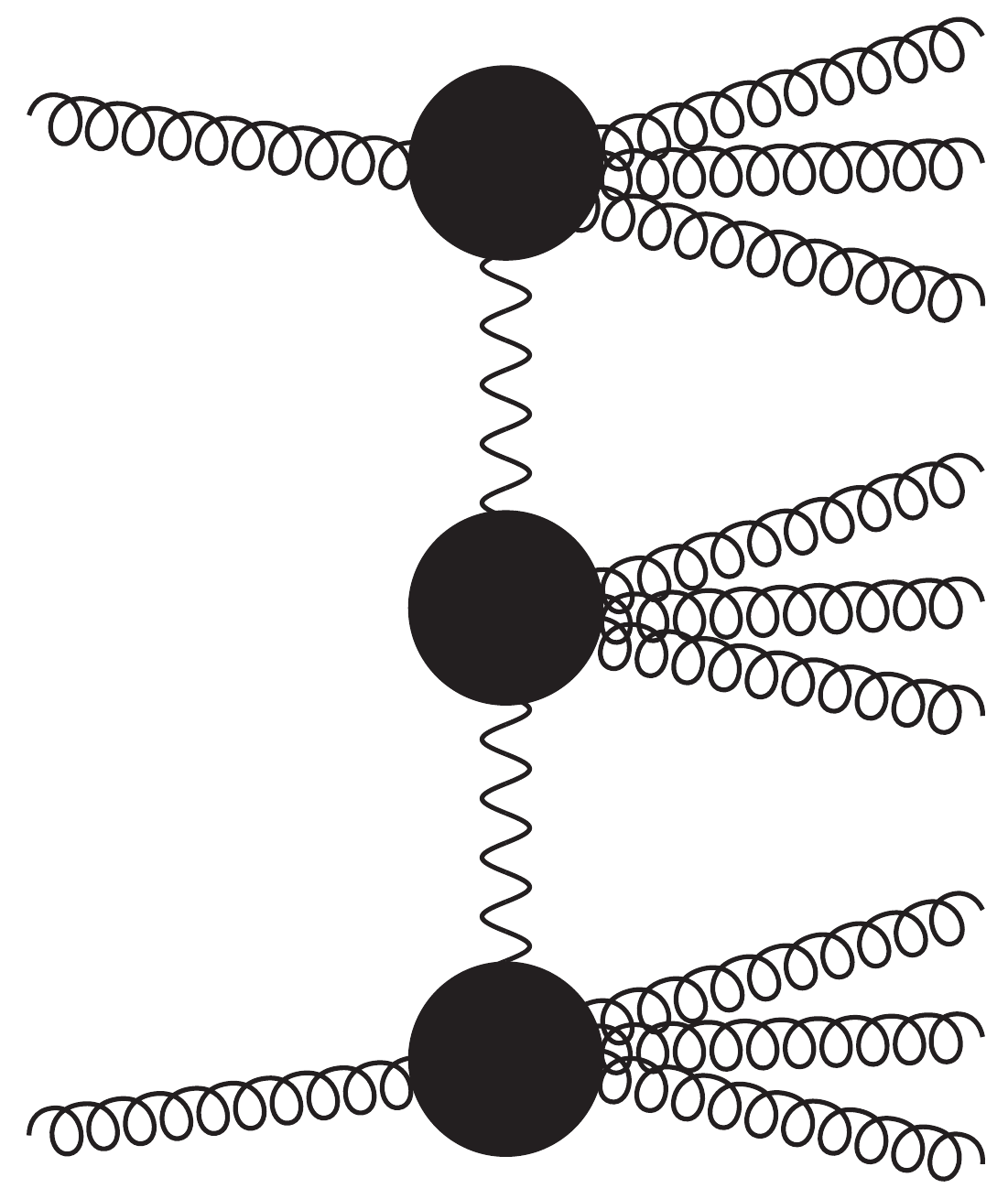}}
  \caption{\small Gluon production in quasi-multi-Regge kinematics. In the effective action framework, each of the clusters separated in rapidity is connected through reggeized gluon exchange.}
  \label{fig:central_production}
\end{figure}

While the effective action can be used to generate the necessary set
of rules to construct the high energy limit of quasi-elastic
amplitudes as discussed in the previous paragraph and depicted in
Fig.~\ref{fig:ggg}, its range of applicability reaches further. It allows to generate a general production amplitude in 
quasi-multi-Regge-kinematics (QMRK), see
Fig.~\ref{fig:central_production}. In this case, besides particle production in the fragmentation region of the 
initial scattering particles, arbitrary many clusters of particles can be produced at central
rapidities, with these clusters being strongly ordered in rapidity among themselves.

From Eq.~\eqref{eq:effac} we notice that the high energy effective
action does not correspond to an effective field theory in the conventional Wilsonian sense.  
Since we now have an enhanced set of Feynman diagrams it is needed to device some 
cut-off or subtraction procedure to avoid double counting in our calculations. In the remaining of this 
review we explain a convenient method to achieve this in two different calculations. 

\section{Forward jet vertex at NLO}
\label{sec:qqNLO}
Let us first use Lipatov's effective action by applying it to the study of the high energy limit of a simple QCD process at NLO: quark-quark scattering.
\subsection{The Born-level result}
\label{sec:tree-level}

\begin{figure}[htb]
  \centering
\parbox{5cm}{\center \includegraphics[height = 2.8cm]{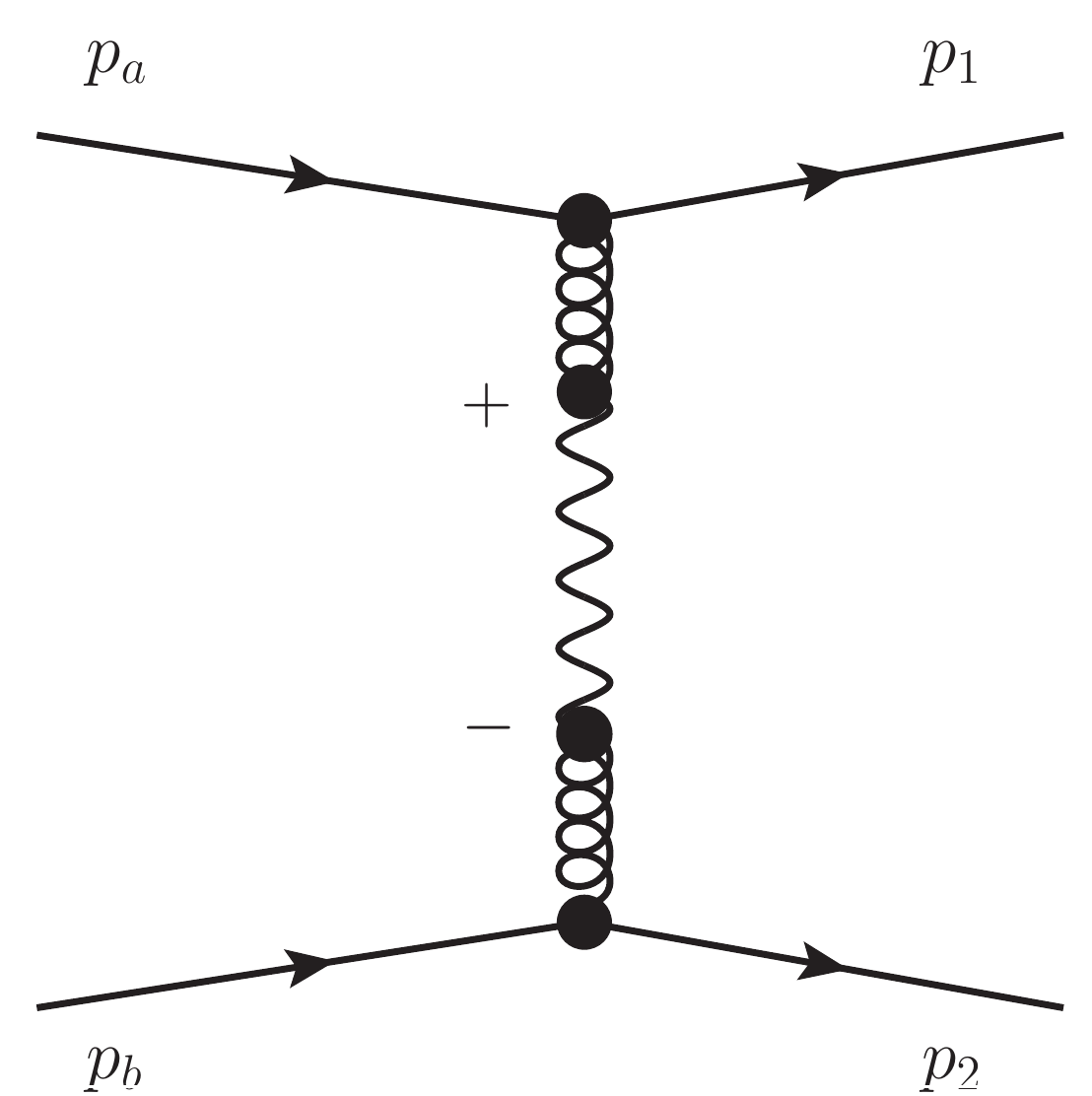}}
\parbox{5cm}{\center \includegraphics[height = 2.8cm]{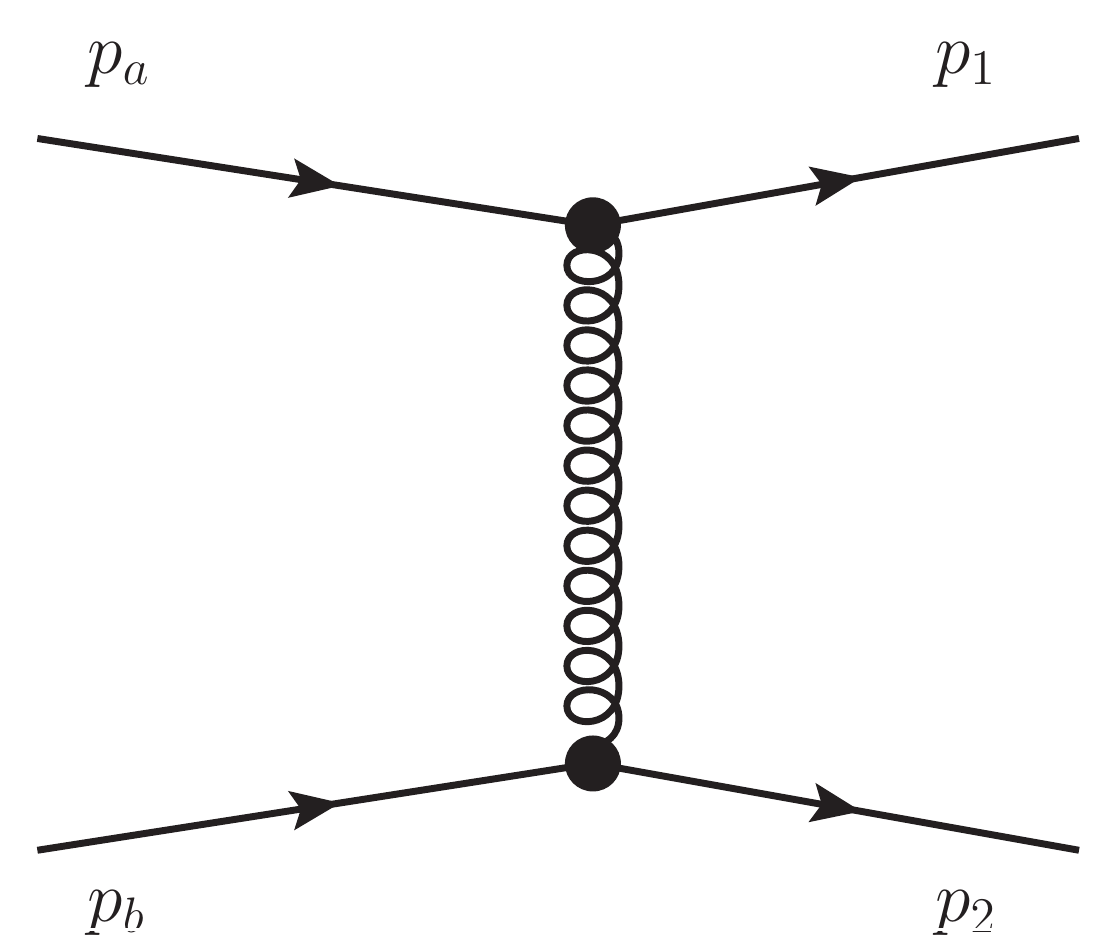}} 

\parbox{5cm}{\center (a)} \parbox{5cm}{\center (b)} 
  \caption{\small Quark-quark scattering mediated by (a): reggeized gluon exchange, (b): gluon exchange.}
  \label{fig:BornLeveldiagrams}
\end{figure}

At Born level the effective action in Eq.~\eqref{eq:effac} provides two Feynman diagrams for the process $p_a + p_b \to p_1 + p_2$, depicted in Fig.~\ref{fig:BornLeveldiagrams}.
To determine the  Born level cross-section in the high energy limit  we need to evaluate diagram Fig.~\ref{fig:BornLeveldiagrams}.a. It  contains the coupling of a reggeized gluon to the on-shell quarks   which reads 
\begin{align}
i\mathcal{M}_{qr^* \to q}^{(0)} & = 
\parbox{2.5cm}{\includegraphics[width = 3cm]{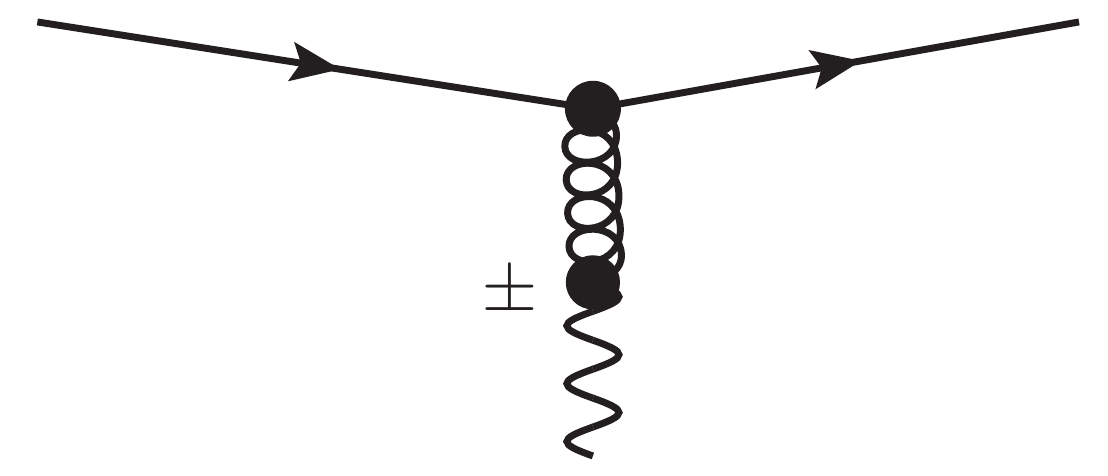}} = 
\bar{u}_{\lambda'}(p')  ig t^a \hspace{-.08cm}\not{\hspace{-.08cm}n}^\pm u_\lambda(p) =i g t^a   2 p^\pm \delta_{\lambda\lambda'}.
\end{align}
Evaluating the differential cross-section 
\begin{align}
  \label{eq:crosssec}
  d \hat{\sigma}^{(0)}_{q_aq_b} =  h^{(0)}_{a}({{\bm k}}) h^{(0)}_{b}({\bm k}) d^{d-2} {\bm k} 
\end{align}
in  $d=4+2 \epsilon$ dimensions with the  $\overline{\rm MS}$ scheme coupling
 $ \alpha_s  = \frac{g^2 \mu^{2\epsilon} \Gamma(1-\epsilon)}{(4 \pi)^{1 + \epsilon}}$
and  $C_F = \frac{N_c^2 -1}{2 N_c}$  we find 
\begin{align}
\overline{| \mathcal{M}^{(0)}|^2}_{qr^* \to q} &=  \frac{1}{4N_c (N_c^2 - 1)}\sum_{\lambda\lambda'}| \mathcal{M}^{(0)}|^2_{qr^* \to q}  =  \frac{ 4 g^2  C_F}{{N_c^2 -1}}   p_a^{+2}, 
\end{align}
and the leading order (LO) quark impact factor $h^{(0)}_{a}({\bm k}) $ in Eq. \eqref{eq:crosssec} is given by
\begin{align}
h_a^{(0)}({\bm k})=\frac{C_F}{\sqrt{N_c^2 -1}} \frac{2^{1+\epsilon}}{\mu^{2\epsilon} \Gamma(1-\epsilon)} \frac{1}{\bm{k}^2},
\end{align}
in agreement with ~\cite{Ciafaloni:1998hu}. The same result can be obtained by evaluating the diagram in Fig.~\ref{fig:BornLeveldiagrams}.b in the limit
$|t|/s \to 0$.  In~\cite{LevSeff} it was suggested to  introduce a factorization parameter $\eta$ 
to associate the high energy region $|t|/s < \eta$ with diagram
Fig.~\ref{fig:BornLeveldiagrams}.a, and the low energy region  $|t|/s > \eta$ with 
Fig.~\ref{fig:BornLeveldiagrams}.b.
Alternatively, it is possible to  subtract the high energy cross-section from  the QCD diagram in Fig.~\ref{fig:BornLeveldiagrams}.b. More precisely,  if the QCD cross-section reads
\begin{align}
  \label{eq:lowener}
   \left( d \hat{\sigma}_{q_a q_b}^{(0)}\right)_{\text{QCD}} & = \frac{1}{2s} \frac{1}{4 N^2 _c}
\sum_{\substack{ \text{spin} \\ \text{color}}}   \left | \parbox{3.5cm}{\center \includegraphics[height = 2.5cm]{4pointLOQCD.pdf}}  \right|^2 d \Phi^{(2)}
\end{align}
where $d \Phi^{(2)}$ denotes the 2 particle Lorentz invariant phase space measure, we can define  a low energy coefficient in the form
\begin{align}
  \label{eq:coeffLO}
   d  \hat{\mathcal{C}}_{q_a q_b}^{(0)} & =     \left( d \hat{\sigma}_{q_a q_b}^{(0)}\right)_{\text{QCD}} 
-
   d \hat{\sigma}_{q_a q_b}^{(0)}.
\end{align}
The complete effective action result then consists of the sum of the high energy
cross-section in Eq.~\eqref{eq:crosssec} and the low energy coefficient in 
Eq.~\eqref{eq:coeffLO} which by construction agrees with the QCD
result.  The leading term of the high energy expansion of the QCD
cross-section is then formally obtained by dropping the low energy
contribution in Eq.~\eqref{eq:coeffLO}.

This procedure can be applied in general to any class of effective
action matrix elements stemming from Eq.~\eqref{eq:effac}. From
those amplitudes with internal QCD propagators only, to which the
reggeized gluon couples as an external (classical) 
field, one subtracts the corresponding high energy factorized
amplitudes with reggeized gluon exchange. The subtracted coefficient
is then local in rapidity. 

\subsection{Real corrections}

The real corrections to the Born level process can be organized into three contributions to the five-point amplitude with central and quasi-elastic gluon production:

\begin{figure}[htb]
  \centering
  \parbox{2.5cm}{\center \includegraphics[width = 2.2cm]{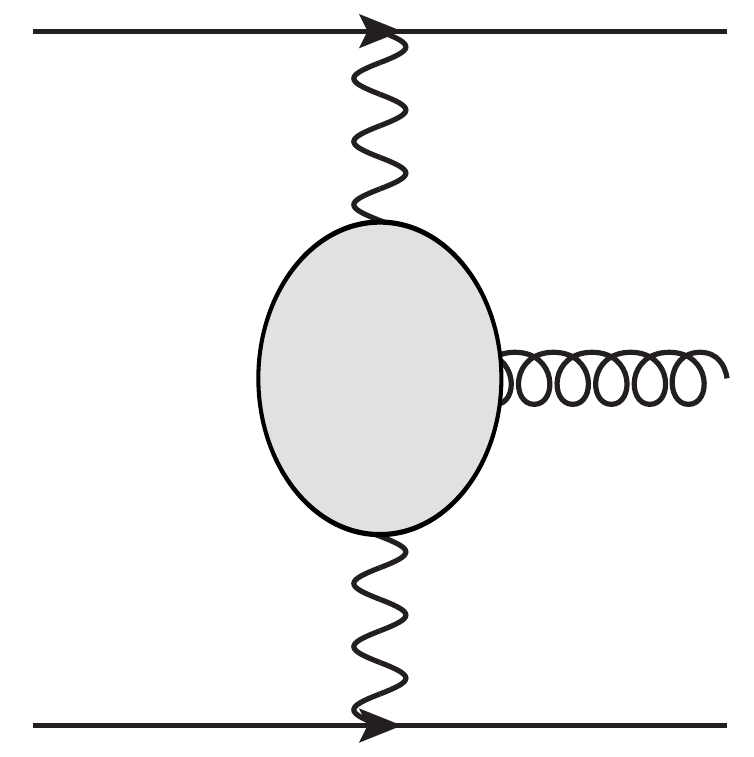}}
 \parbox{2.5cm}{\center \includegraphics[width = 2.2cm]{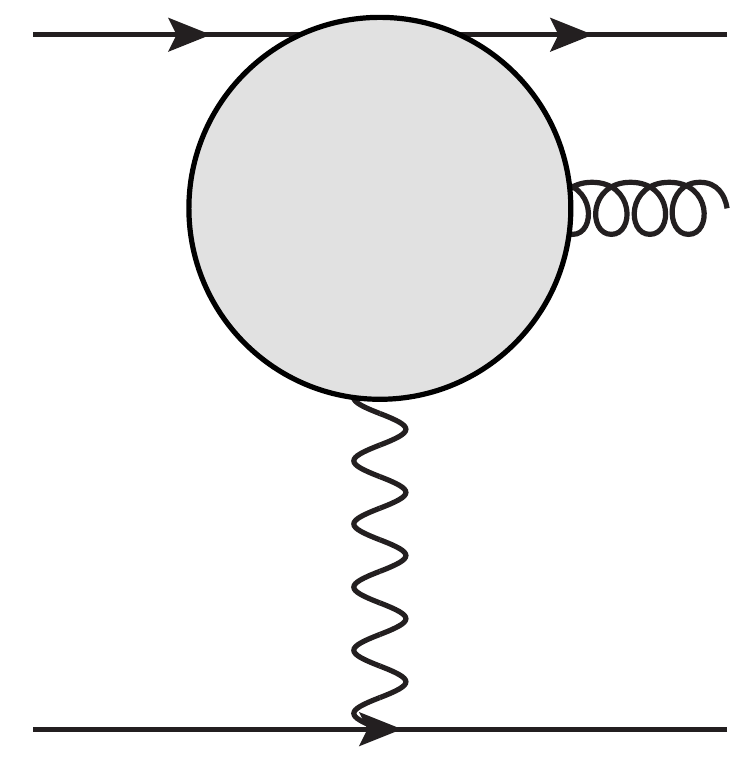}}
 \parbox{2.5cm}{\center \includegraphics[width = 2.2cm]{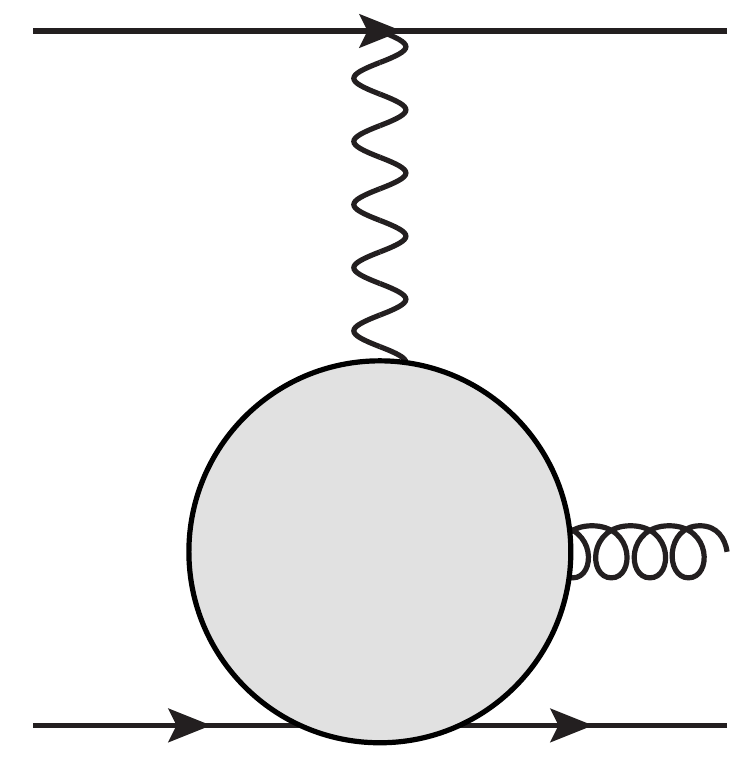}}
\end{figure}
Integrating over the longitudinal   phase space of the produced gluon we will find  divergences  which can be regularized using an
explicit cut-off in rapidity.  
The central production amplitude,  obtained from the sum of 
the following three effective diagrams, 
yields the unintegrated real part of the forward  LO  BFKL kernel:
\begin{eqnarray}
   \parbox{2cm}{\center \includegraphics[width=1.2cm]{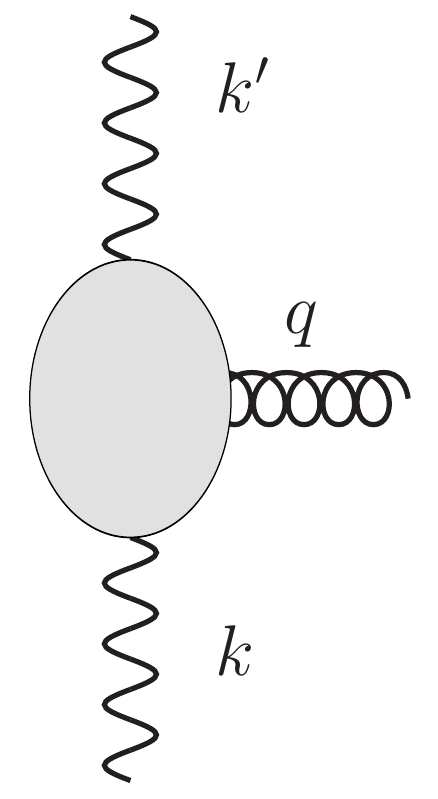}} 
& =&
   \parbox{2cm}{\center \includegraphics[width=1.2cm]{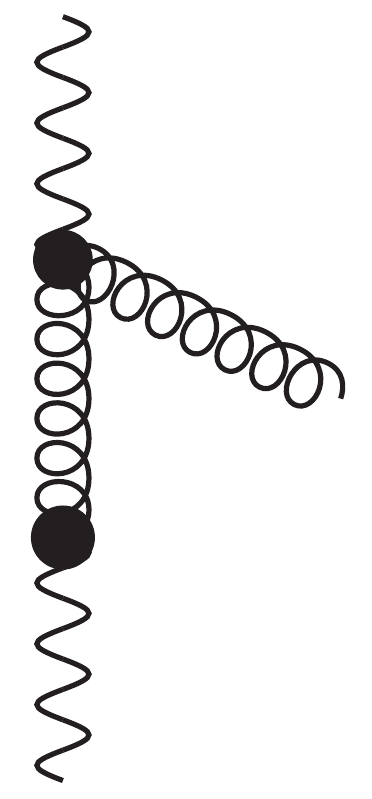}} +
   \parbox{2cm}{\center \includegraphics[width=1.2cm]{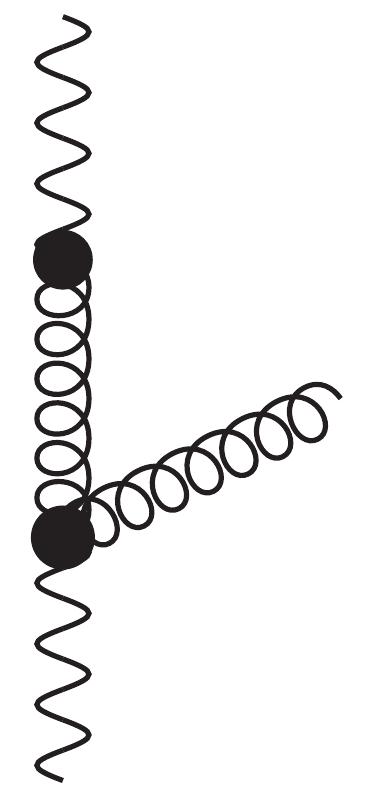}} +
   \parbox{2cm}{\center \includegraphics[width=1.2cm]{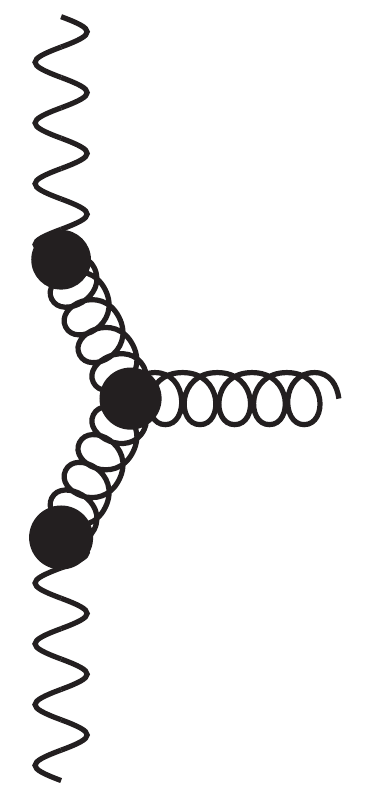}}, \nonumber
\end{eqnarray}
with the  following Sudakov decomposition of the momenta 
\begin{align}
  \label{eq:sudakov}
   k'  = \frac{q^+}{2} n^+  + 
{\bm k}',\qquad
 k  = \frac{k^-}{2} n^+ + \bm {k},\qquad
  q  = \frac{q^+}{2} n^+ +
\frac{k^-}{2} n^+  + {\bm q},
\end{align}
with
\begin{align}
  \label{eq:sduwith}
  k^- & =  \frac{{\bm q}^2}{q^+} & {\bm k}' & = {\bm q} - {\bm k}.
\end{align}
Note that the condition ${k'}^- = 0 = k^+$ appears as a  direct consequence of the kinematic
constraint in Eq.~\eqref{eq:kinematic}. The squared amplitude, averaged
over color of the incoming reggeized gluons and summed over final state color
and helicities reads
\begin{eqnarray} 
  \overline{|\mathcal{M}|^2}_{r^*r^* \to g} &=& \frac{16g^2 N_c}{N_c^2 - 1} 
         \frac{{\bm k'}^2 {\bm k}^2 }{ {\bm q}^2}.
\end{eqnarray}
It leads to the following production vertex
\begin{eqnarray}
  V({\bm q}; {\bm k}, {\bm k}') &=& 
  \frac{N_c^2 - 1}{8 (2 \pi)^{3+2 \epsilon}  {\bm k}^2  {\bm k'}^2   }
  \overline{|\mathcal{M}|^2}_{r^*r^* \to g} ~=~   \frac{\alpha_s N_c}{\mu^{2 \epsilon} \pi^{2 + \epsilon} {\bm q}^2} .
\end{eqnarray}
The exclusive differential cross section for central production then reads
\begin{align}
\label{eq:cent_prod}
 d \hat{\sigma}^{(c)}_{ab} &=  h^{(0)}_a( {\bm k'}) h^{(0)}_b({ \bm k})    
 \mathcal{V}({\bm q}; { \bm k}, {\bm k'}, \eta_a, \eta_b)  d^{2+2\epsilon} {\bm k'} d^{2+2\epsilon} {\bm k} \,d \eta,
 \end{align}
 where ${\cal V}({\bm q }; {\bm k}, {\bm k'}, \eta_a, \eta_b) \equiv
 V({\bm q}; {\bm k}, {\bm k'}) \theta(\eta_a - \eta)\theta(\eta -
 \eta_b)$ corresponds to the regularized production vertex with cut-offs
 $\eta_{a,b}$, to be evaluated in the limit $\eta_{a,b} \to
 \infty$. Once integrated over the full range in $\eta$,
 Eq.~\eqref{eq:cent_prod} without regulators would result into a
 (longitudinal) high energy divergence, proportional to the real part
 of the LO BFKL kernel. \\
 
 For the quasi-elastic contribution $q(p_a) r^*(k) \to g(q)q(p)$ we first evaluate the sum of the effective
 diagrams
\begin{eqnarray}
 \parbox{4cm}{\center \includegraphics[width=3.5cm]{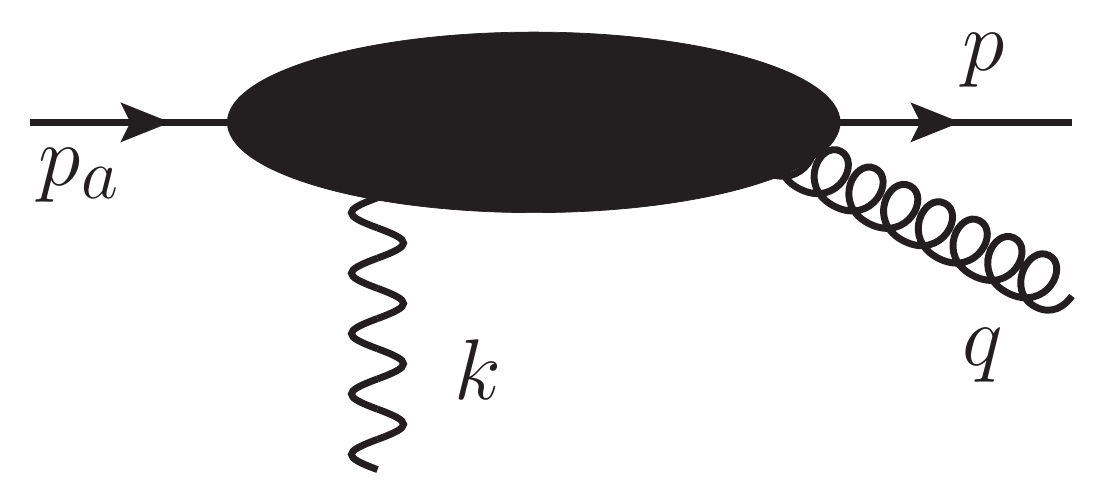}} \hspace{-0.5cm}&=&
\hspace{-0.6cm}  \parbox{4cm}{\center \includegraphics[width=3.5cm]{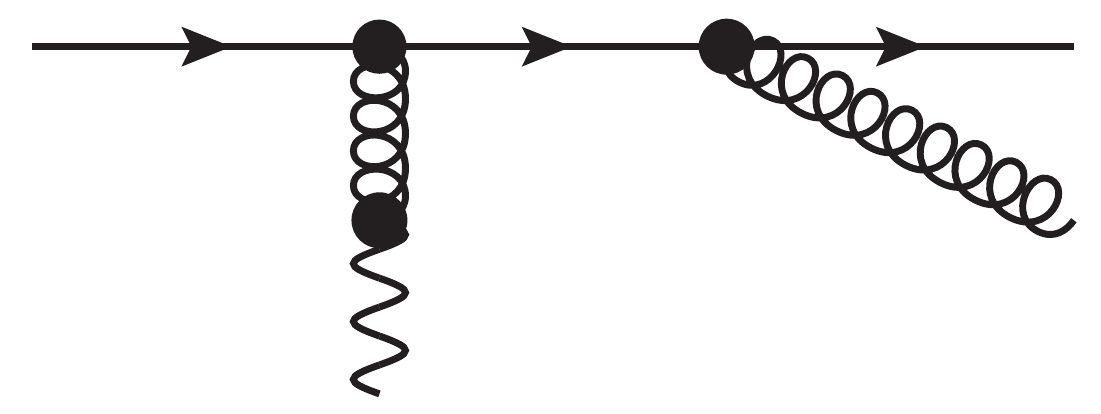}} \hspace{-0.3cm}+
\hspace{-0.3cm}   \parbox{4cm}{\center \includegraphics[width=3.5cm]{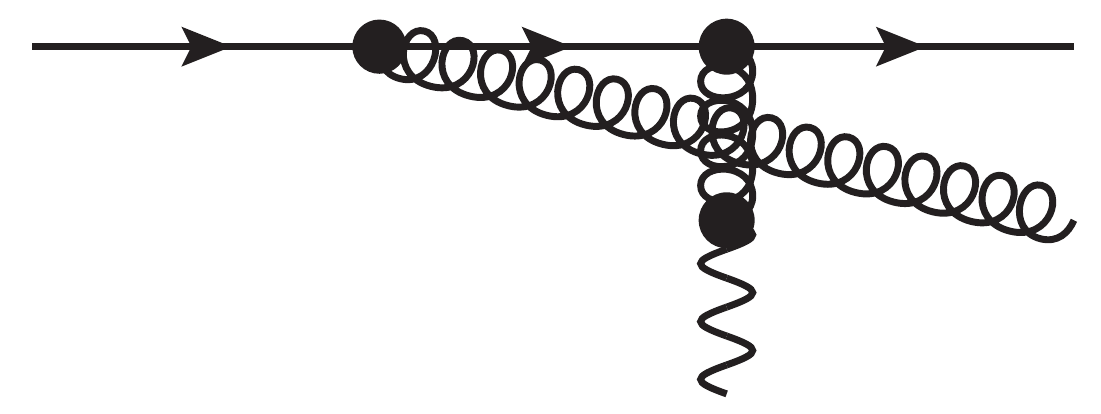}} \hspace{-0.3cm} \nonumber \\
&+&
\hspace{-0.3cm} \parbox{4cm}{\center \includegraphics[width=3.5cm]{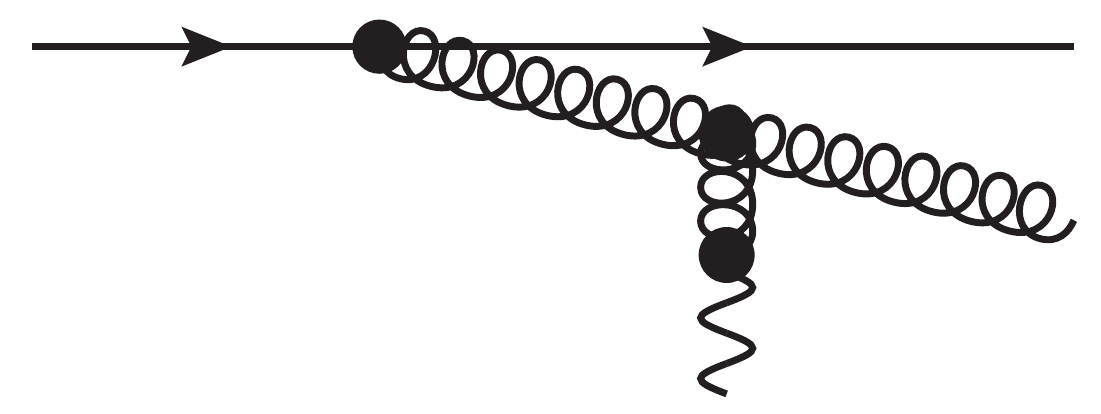}} \hspace{-0.3cm} +
\hspace{-0.3cm} \parbox{4cm}{\center \includegraphics[width=3.5cm]{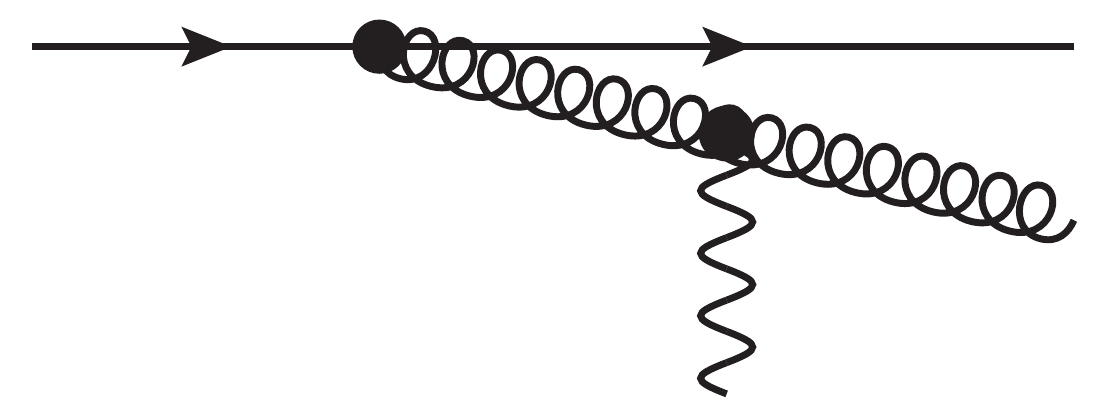}} .\nonumber
\end{eqnarray}
With the Sudakov decomposition of external momenta given by
\begin{align}
  \label{eq:suda_quasi}
 p_a & = \frac{p_a^+}{2} n^- &  
 p & = \frac{(1-z)p_a^+}{2} n^-  + \frac{p^-}{2} n^+ + {\bm k}  - {\bm q}, 
\notag \\
k & = \frac{k^-}{2}n^+ + {\bm k}
&
q &= \frac{z p_a^+}{2} n^- + \frac{k^- - p^-}{2} n^+ + {\bm q},
\end{align}
  with $z= q^+ / p^+_a$ and
\begin{align}
  \label{eq:withquasi}
  p^- & = \frac{({\bm k} - {\bm q}^2)}{(1-z) p_a^+}, & k^- & = \frac{{\bm \Delta}^2 + z(1-z){\bm k}^2}{(1-z) p_a^+}, & {\bm \Delta} &= {\bm q} - z {\bm k}.
\end{align}
The squared amplitude reads
\begin{eqnarray}
\overline{|\mathcal{M}|^2}_{r^*q \to qg}& =& 
\frac{8 g^4  p_a^{+2}}{{N_c^2 -1}} \frac{\mathcal{P}_{gq}(z,\epsilon)}{\bm{\Delta}^2 \bm{q}^2}  \frac{(1-z)z \bm{k}^2}{\bm{k}'^2} 
\theta\left(z - e^{-\eta_b}\frac{\sqrt{q_\perp^2}}{p_a^+} \right)
 \nonumber \\ && \qquad \qquad \qquad \qquad \cdot
\left[C_F z^2 {k}'^2 + N_c (1-z) \bm{\Delta}\cdot \bm{q} \right]
,
\end{eqnarray}
where $\mathcal{P}_{gq}(z,\epsilon) = C_F \frac{1 + (1-z)^2 + \epsilon z^2}{z}$ is the real part of the $q \to g$ splitting function and  ${\bm k}' = {\bm q} - {\bm k}$. The lower cut-off   $\eta_b$ on the rapidity  $\eta = \ln {\bm q}^2/ (z p_a^+)$ of the gluon has been introduced, in direct analogy to the corresponding lower cut-off for the central production vertex. The upper limit for the quasi-elastic contribution is bounded by kinematics 
since $z < 1$.  Putting together these results, the real corrections to the jet vertex are 
\begin{eqnarray}
h^{(1)}({\bm k}) dz d^{2+2\epsilon}{\bm q}  &=&  \frac{\sqrt{N_c^2 - 1}}{(2 p_a^+)^2} 
\int \frac{d k^-}{(2 \pi)^{2 + \epsilon}}  d \Phi^{(2)} | \mathcal{M}|^2_{qg^* \to qg} \frac{1}{{ \bm k}^2},
\end{eqnarray}
with the two-particle phase space explicitly given by
\begin{eqnarray}
d\Phi^{(2)} &=&  \frac{1}{2p_a^+ (2\pi)^{2 + 2\epsilon}} dz d^{2+2\epsilon}{\bm q } 
\frac{1}{(1-z)z} \delta\left(k^- - \frac{{ \bm \Delta}^2 + z (1-z) { \bm k}^2}{(1-z)p_a^+}\right).
\end{eqnarray}
The final result exactly agrees with the equivalent one in~\cite{Ciafaloni:1998hu}:
\begin{align}
h^{(1)} (z; \,&  {\bm k}^2, {\bm k'}^2, {\bm q}^2)  = h^{(0)} ({\bm k}^2) \, \cdot \, 
\mathcal{F}_{qqg}(\bm{q},\bm{k},z)
 \\
\mathcal{F}_{qqg}(\bm{q},\bm{k},z)&=
 \frac{\alpha_s}{2\pi} \frac{\mathcal{P}_{gq}(z,\epsilon)}{\pi_\epsilon} \frac{1}{{\bm q}^2 {\bm \Delta}^2} 
\theta \left(z - e^{-\eta_b}\frac{\sqrt{{\bm q}^2}}{p_a^+} \right)
\left[C_F z^2 {\bm k'}^2 + N_c (1-z) {\bm \Delta}\cdot {\bm q} \right] 
.\notag
\end{align}
Note that in the limit $z \to 0$, which corresponds to a large distance in rapidity between the final state quark and gluon, the above expression turns into the central production vertex, multiplied by the leading order impact factor
\begin{align}
  \label{eq:limit}
  \lim_{z \to 0} h^{(1)} (z; {\bm k}^2, {\bm k'}^2, {\bm q}^2) dz   &= h^{(0)} ({\bm k}^2) V({\bm q}; {\bm k}, {\bm k'}) d \eta.
\end{align}
Simply adding quasi-elastic and central production cross-section 
therefore leads to the expected over-counting. For production processes at 
tree-level, there are two immediate 
solutions. A physical intuitive solution slices the longitudinal phase space and associates a fixed  range of rapidity
 to gluon production in the quasi-elastic (where the gluon separation of gluon and one final state quark is small) and central region (with a significant
separation between both quarks). Here we choose to follow a different treatment
suggested at the end of Sec.~\ref{sec:tree-level} and 
subtract  the central production contribution of 
Eq.~\eqref{eq:limit}, including the full cut-off dependence as given
in Eq.~\eqref{eq:cent_prod} from the quasi-elastic correction, {\it i.e.}, schematically
\begin{eqnarray}
   \parbox{4cm}{\center \includegraphics[height=1.3cm]{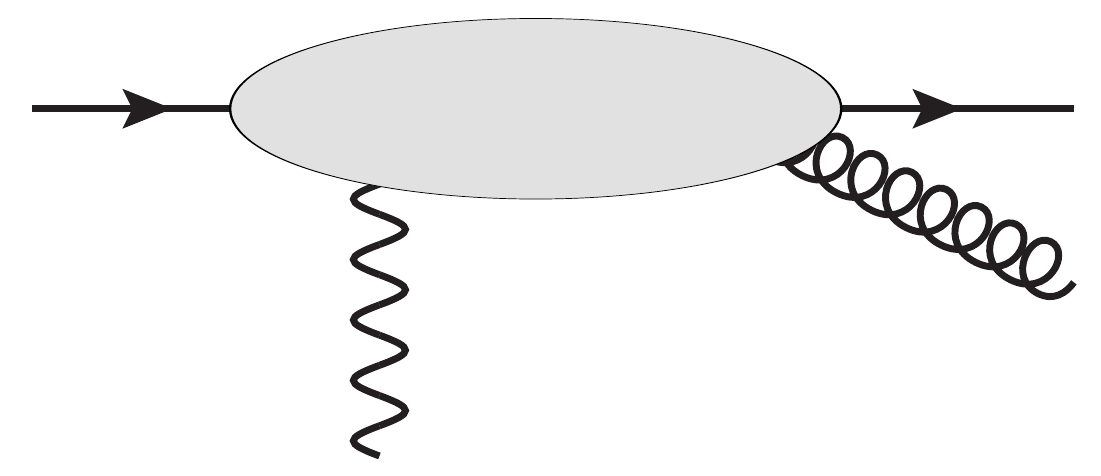}} &=&
\parbox{4cm}{\center \includegraphics[height=1.3cm]{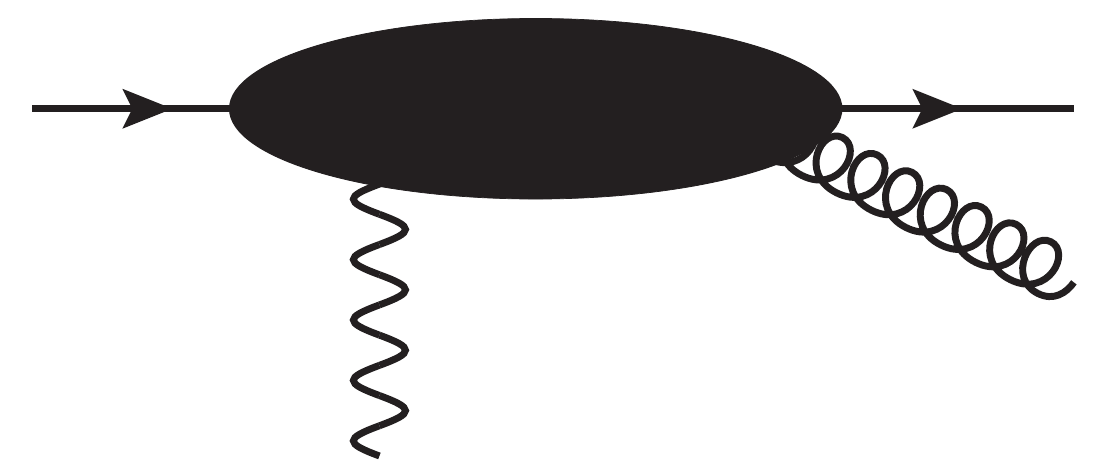}}  - \parbox{4cm}{\center \includegraphics[height=1.3cm]{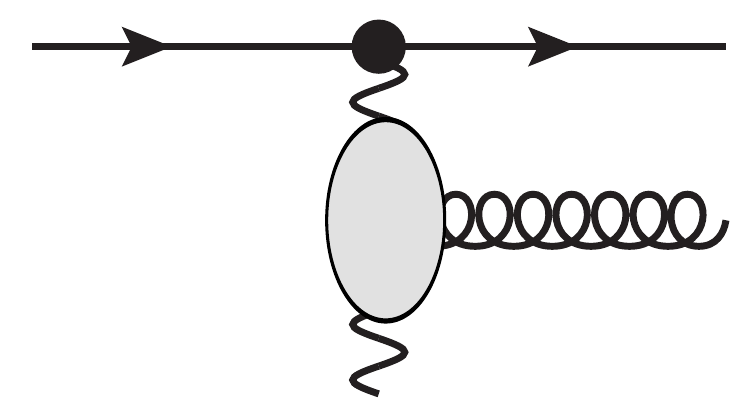}}. \nonumber
\end{eqnarray}
In this way the  subtracted quasi-elastic contribution to the exclusive differential cross-section is already local in rapidity space, in the sense that it only depends on the regulator $\eta_a$ 
\begin{eqnarray}
d \hat{\sigma}^{(qea)}_{ab} &=& h^{(0)}_a (\bm{k}') h^{(0)}_b(\bm{k})  
\mathcal{G}^a_{qqg}(\bm{k}, \bm{q},z, \eta_a, \eta_b )  d^{2+2\epsilon}\bm{q} \,d^{2+2\epsilon} \bm{k}\, dz, 
\end{eqnarray}
where
\begin{align}
  \label{eq:limit}
  \mathcal{G}^{a,b}_{qqg} & = \lim_{\eta_b \to \infty} \left[ \mathcal{F}_{qqg} - \mathcal{V}({\bm q}; {\bm k}, {\bm k'}, \eta_a, \eta_b) \right]
=
\frac{\alpha_s}{2\pi}  \bigg\{ \frac{\mathcal{P}_{gq}(z,\epsilon)}{\pi^{1 + \epsilon} \Gamma(1-\epsilon)} 
\bigg[ \frac{ C_F z^2 { \bm  k}^{'2}}{{ \bm q}^2 { \bm  \Delta}^2}
\notag \\
& 
+ \frac{N_c (1-z) {\bm \Delta} \cdot {\bm q}}{{ \bm q}^2 { \bm \Delta}^2}\bigg]
 - \frac{N_c}{z} \frac{1}{ {\bm q}^2}  \bigg\} 
+\frac{N_c}{z} \frac{1}{{\bm q}^2}
\theta\left(z - e^{\pm \eta_{a,b}}\frac{\sqrt{{\bm q}^2}}{p_a^+} \right).
\end{align}
The complete exclusive differential cross-section is then given as the sum of central and quasi-elastic contributions,
\begin{eqnarray}
  d \hat{\sigma}_{ab} &=&  d \hat{\sigma}^{(c)}_{ab} +  d \hat{\sigma}^{(qea)}_{ab} +  d \hat{\sigma}^{(qeb)}_{ab},
\end{eqnarray}
for which the dependence on the regulators $\eta_{a,b}$ cancels. Let us now consider the corresponding 
virtual contributions.

\subsection{Virtual corrections: pole prescription and regularization}
\label{sec:virtual-corrections}

To evaluate loop corrections within the effective action it is
necessary to fix a prescription for the light-cone pole of the
induced vertex in Fig.~\ref{fig:feynrules0p2}.c.  Although 
the final result for a scattering amplitude must be independent
from the chosen prescription, a convenient one can
considerable help simplify the full calculation.  In the
following we use the prescription suggested in~\cite{Hentschinski:2011xg}. 
We replace the unregulated operator
$W_\pm(v)$ in the induced Lagrangian of Eq.~\eqref{eq:1efflagrangian} by the expression
\begin{align}
  \label{eq:Wreg}
  W^\epsilon_\pm [v] & =  \frac{1}{2} \left[ \mathcal{P}_{\text{A}} \left( v_\pm \frac{1}{D_\pm - \epsilon} \partial_\pm \right)  + \mathcal{P}_{\text{A}} \left( v_\pm \frac{1}{D_\pm + \epsilon} \partial_\pm \right) \right], 
\end{align}
  The projector $\mathcal{P}_{\text{A}}$ is needed since the
color structure of the induced vertices is defined in terms of only
antisymmetric color structure, in terms of the 
SU$(N_c)$ structure constants $f^{abc}$. While for non-zero values of
the operators $1/\partial_\pm$ this happens automatically, a pole
prescription of the above kind leads to momentum space expressions
which are proportional to symmetric color tensors multiplied by a
delta-function in one of the light-cone momenta.  We remove these subleading terms using 
a suitable projector, keeping in this way the same color structure as in the unregulated case.
The projector $\mathcal{P}_{\text{A}}$ acts then order by order in $g$
on the SU$(N_c)$ color structure of the gluonic fields $v_\pm(x) = -
it^a v^a (x)$,
\begin{align}
  \label{eq:Wreg2}
  \mathcal{P}_{\text{A}} \bigg(& v_\pm \frac{1}{D_\pm - \epsilon} \partial_\pm \bigg)  \equiv
 -i\bigg(  P^{(1)}_{\text{A}}(t^a)  v^a_\pm  - (-ig) v^{a_1}_\pm \frac{1}{\partial_\pm - \epsilon}  v^{a_2}_\pm  P^{(2)}_{\text{A}}\left(t^{a_1}t^{a_2} \right)
\notag \\
& \qquad 
 +   (-ig)^2 v^{a_1}_\pm \frac{1}{\partial_\pm - \epsilon}  v^{a_2}_\pm  \frac{1}{\partial_\pm - \epsilon}  v^{a_3}_\pm     P^{(3)}_{\text{A}}\left(t^{a_1}t^{a_2}t^{a_3} \right)  - \ldots  \bigg),
\end{align}
where $P_{\text{A}}^{(n)}$ are the projectors of the color tensors
with $n$ adjoint indices on the maximal antisymmetric subsector of
order $n$. The latter can be defined by an iterative procedure,
outlined in~\cite{Hentschinski:2011xg}.  For the $\mathcal{O}(g)$
induced vertex, to which we can restrict in the following, this means
discarding the symmetric color tensor $d^{abc}$ from the
$\mathcal{O}(g)$ induced vertex which arises from Eq.~\eqref{eq:Wreg}
before projecting, resulting into a Cauchy principal value
prescription for the pole in Fig.~\ref{fig:feynrules0p2}.c.

Furthermore, in full analogy with the evaluation of  real corrections,   
 loop diagrams of the effective action  lead  to a new type of
longitudinal divergences, which are not present in conventional quantum corrections to QCD
amplitudes.  For loop calculations  a convenient way to regularize these divergences is to 
introduce an external parameter $\rho$, evaluated in the limit $\rho \to \infty$, which can be interpreted as $\log s$. It  deforms the light-cone four vectors
of the effective action in the form
\begin{eqnarray}
  \label{eq:n+-}
  n^- \to n_a &=& e^{-\rho} n^+ + n^- , \,\,\, \,\,\, n^+ \to n_b ~=~  n^+ + e^{-\rho}   n^-.
\end{eqnarray}

\begin{figure}[htb]
  \centering
  \parbox{1.5cm}{\vspace{0.1cm} \includegraphics[height = 2.5cm]{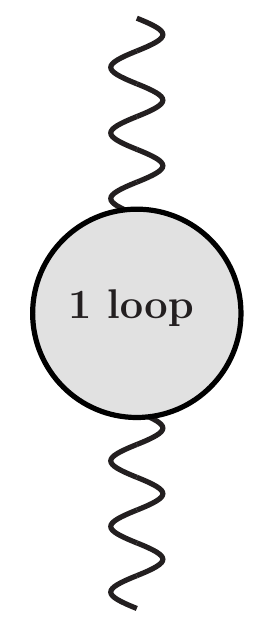}}
= 
    \parbox{1cm}{\vspace{0.1cm} \includegraphics[height = 2.5cm]{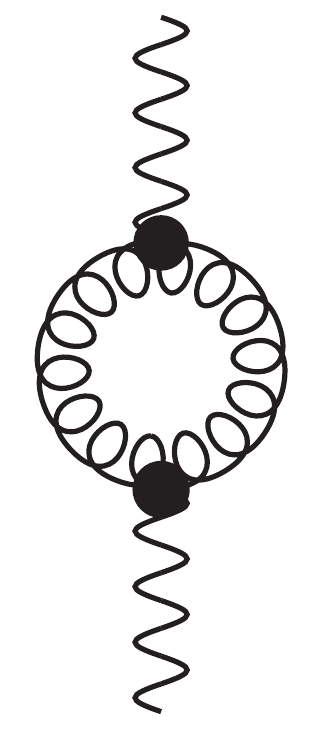}}
  + 
  \parbox{1cm}{\vspace{0.1cm} \includegraphics[height = 2.5cm]{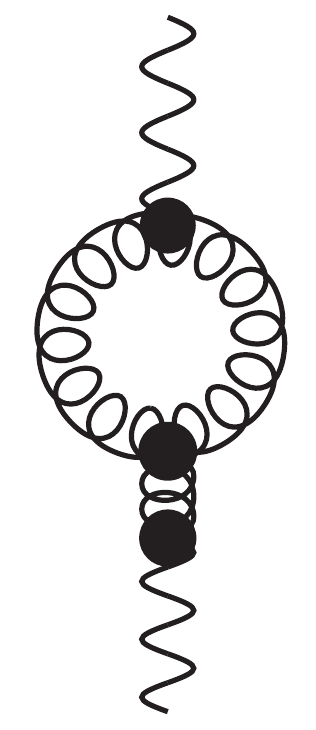}} 
 +
  \parbox{1cm}{\vspace{0.1cm} \includegraphics[height = 2.5cm]{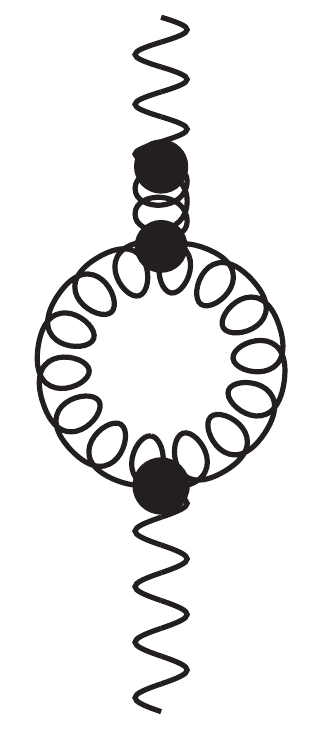}} 
 +
  \parbox{1cm}{\vspace{0.1cm} \includegraphics[height = 2.5cm]{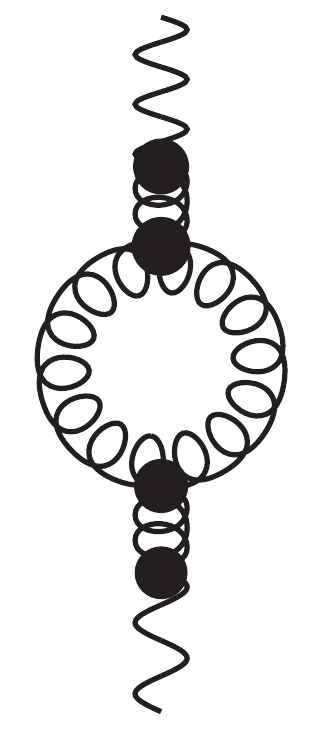}} 
 +
  \parbox{1cm}{\vspace{0.1cm} \includegraphics[height = 2.5cm]{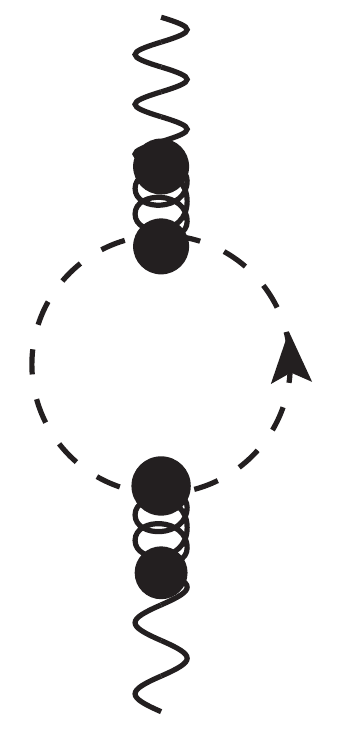}}
+
  \parbox{1cm}{\vspace{0.1cm} \includegraphics[height = 2.5cm]{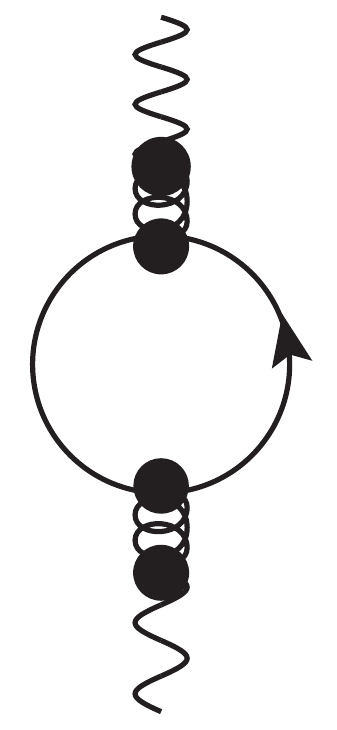}}
\caption{\small Diagrams contributing to the one-loop reggeized gluon self-energy.}
\label{fig:self_1loop}
\end{figure}

 To 
study the virtual corrections it is first needed to obtain the one-loop self energy corrections to the reggeized gluon 
propagator. The contributing diagrams (including ghost loops) are shown in Fig.~\ref{fig:self_1loop}.

\begin{figure}[htb]
  \centering
  \parbox{2cm}{\includegraphics[width = 2cm]{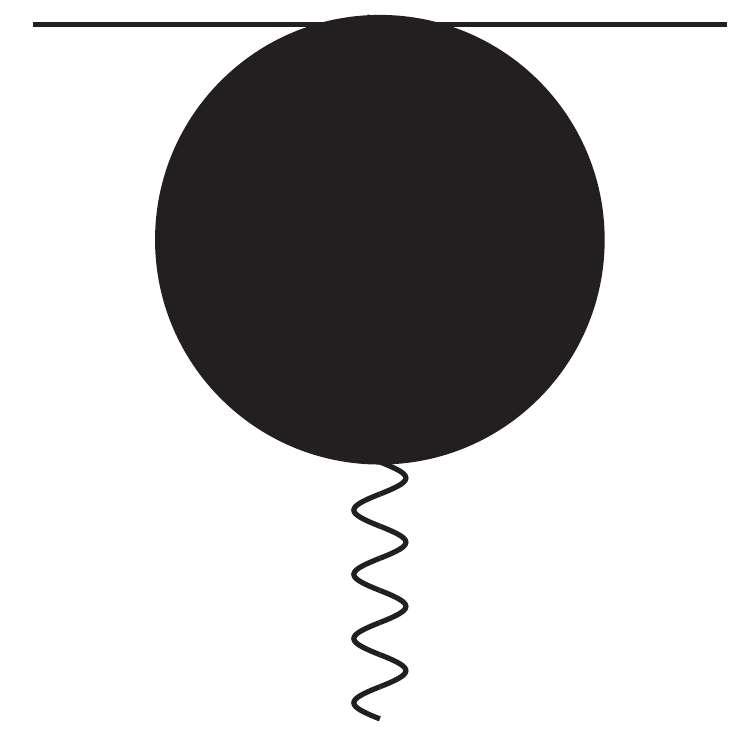}}   =
\parbox{2cm}{\includegraphics[width = 2cm]{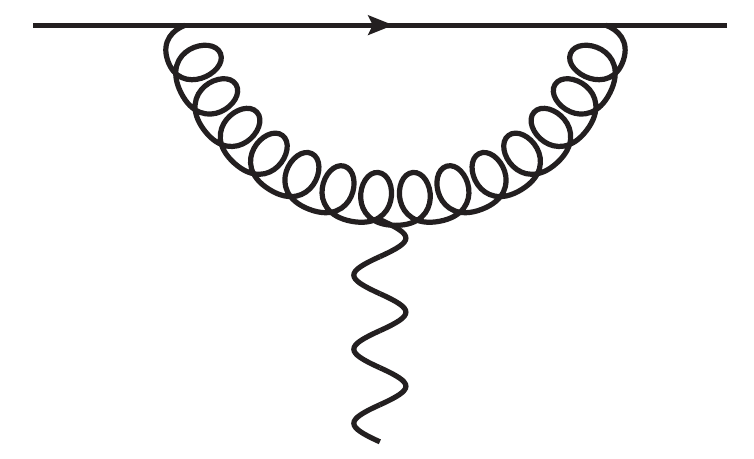}}
  + 
 \parbox{2cm}{\includegraphics[width = 2cm]{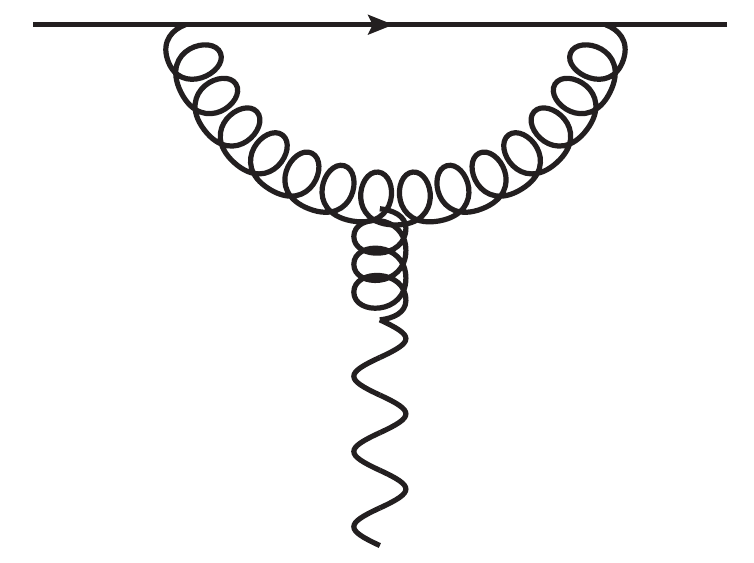}}
  +
\parbox{2cm}{\includegraphics[width = 2cm]{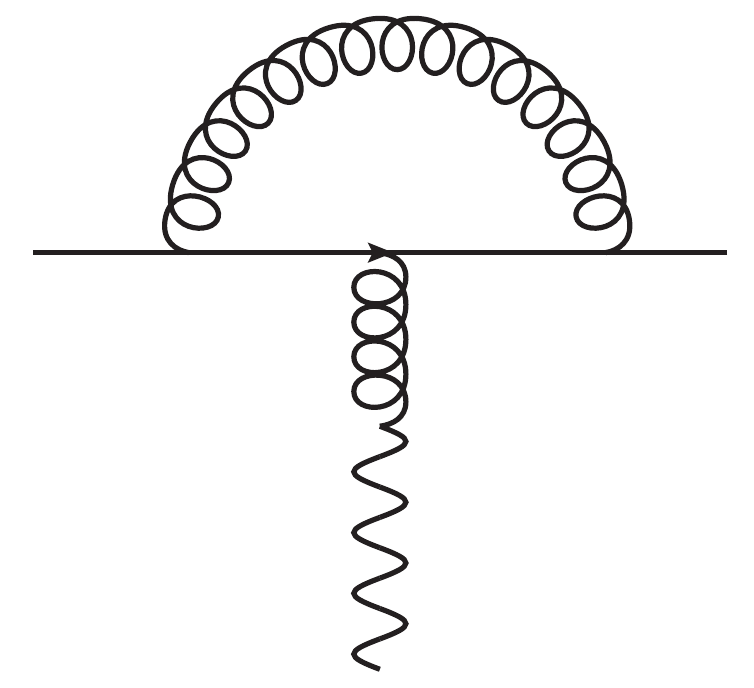}} 
+ \parbox{2cm}{\includegraphics[width = 2cm]{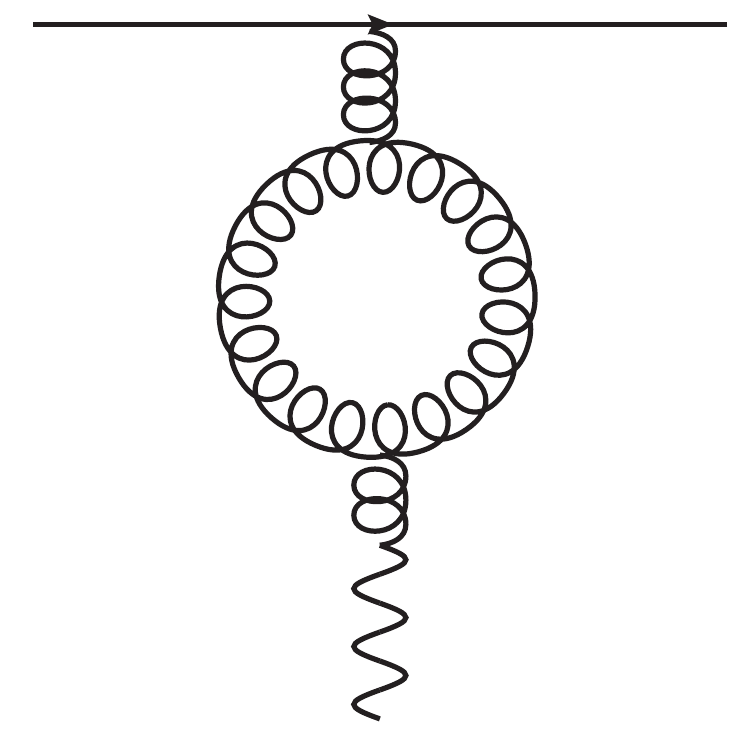}} 
+
\parbox{2cm}{\includegraphics[width = 2cm]{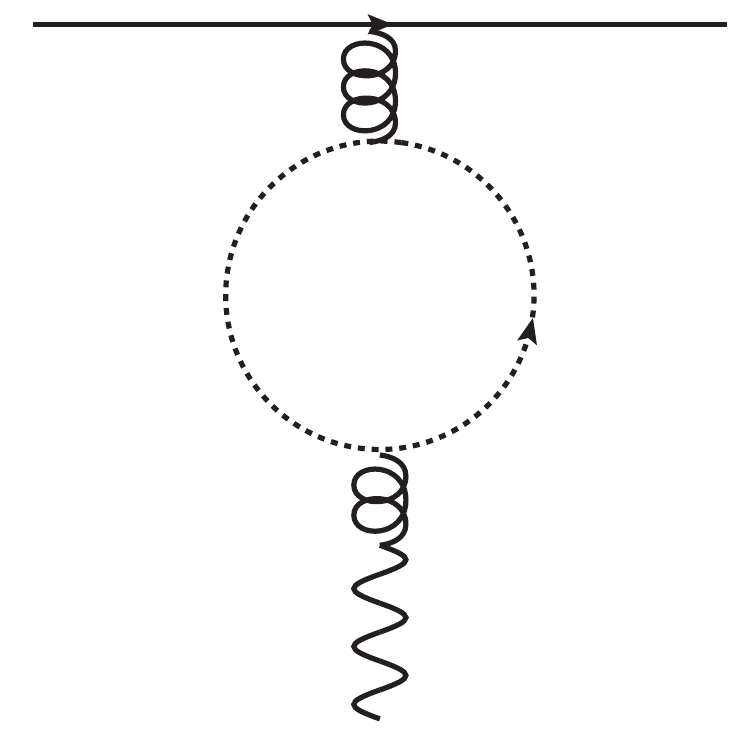}}
+
\parbox{2cm}{\includegraphics[width = 2cm]{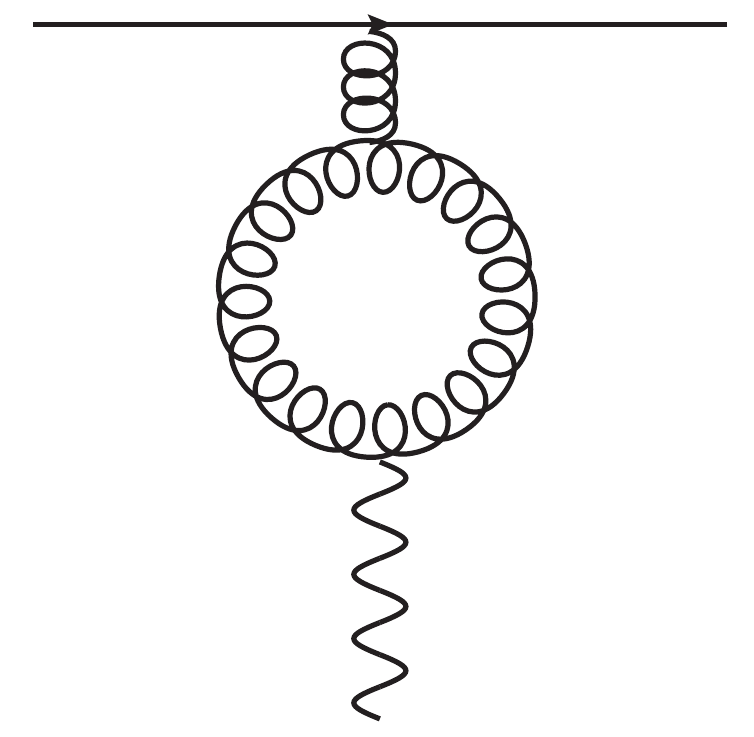}}
+
\parbox{2cm}{\includegraphics[width = 2cm]{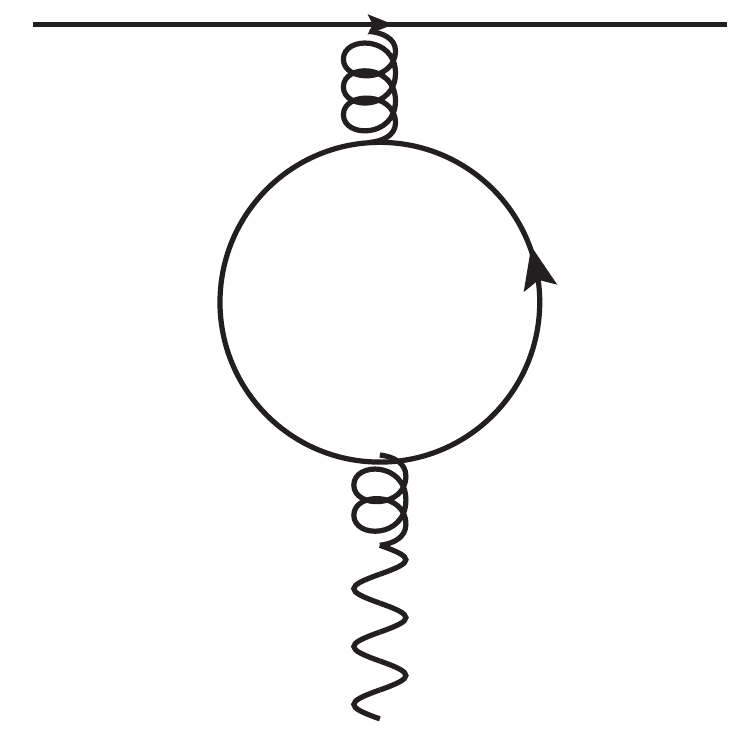}} 

  \caption{\small Contributions to the unsubtracted one-loop quark-quark-reggeized gluon vertex.}
  \label{fig:quasi_graphs}
\end{figure}

Keeping the ${\cal O} (\rho, \rho^0)$ terms, for $\rho  \to \infty$, 
we have the following result:
\begin{align}
\label{eq:self_1loop}
 &  \parbox{1cm}{\includegraphics[width = 1cm]{self_1loop.pdf}} 
=    \Sigma^{(1)}\left(\rho; \epsilon, \frac{{\bm q}^2}{\mu^2}    \right)    
 =
 \frac{(-2i {\bm q}^2) \alpha_s N_c \Gamma^2(1 + \epsilon)}{4 \pi \Gamma(1 + 2 \epsilon)} \left(\frac{{\bm q}^2}{\mu^2} \right)^\epsilon   
\notag \\
& \hspace{1.9cm}
\times 
  \bigg\{  \frac{ i\pi - 2 \rho}{\epsilon}         
- \frac{1}{(1 + 2 \epsilon)\epsilon} \bigg[\frac{5 + 3\epsilon}{3 + 2 \epsilon} 
-\frac{n_f}{N_c}  \left(\frac{2 + 2\epsilon}{3 + 2\epsilon}\right)\bigg] \bigg\}. 
\end{align}
It contains both finite $\sim \rho^0$ and divergent terms $\sim \rho$, where the latter are found to be proportional to the one-loop gluon Regge trajectory. 
The one-loop corrections to the quark-quark-reggeized gluon vertex $ i \mathcal{M}^{(1)}_{qr^* \to q}$ are on the other hand obtained from evaluating the set of diagrams shown in Fig.~\ref{fig:quasi_graphs}.
\\
We obtain the following result 
\begin{align}
  \parbox{2cm}{\includegraphics[width = 2cm]{impaamp.pdf}}
&= {i \mathcal{M}^{(0)}_{qr^*\to  q}}  \frac{\alpha_s N_c}{ 4 \pi \epsilon}  \frac{\Gamma(1 + \epsilon)^2}{\Gamma(1 + 2 \epsilon)}
\left( \frac{{\bm k}^2}{\mu^2} \right)^\epsilon
\bigg[
-  
\left( \ln \frac{-p_a^+}{\sqrt{{\bm k}^2}} +   \ln \frac{p_a^+}{\sqrt{{\bm k}^2}} + {\rho} \right) 
 \notag \\
&   +
 \frac{3 + 2 \epsilon}{N_c^2 2\epsilon (1 + 2 \epsilon)}   -  \frac{11 + 7 \epsilon}{(3 + 2\epsilon)(1 + 2 \epsilon)}
 +  \frac{n_f}{N_c} \frac{2 + 2\epsilon}{(3 + 2\epsilon)(1 + 2 \epsilon)}  
  \notag \\
&
+  \frac{2 + 11 \epsilon}{2 \epsilon(1 + 2 \epsilon)}
 - 
\bigg(
\psi(1-\epsilon) - 2 \psi(\epsilon) + \psi(1)
\bigg)
 \bigg],
\end{align}
where $ \psi(z) = {\Gamma'(z)}/{\Gamma(z)}$.  In analogy with our treatment of the  real NLO corrections we  subtract from the above
result the non-local contributions stemming from the one-loop
corrections to the reggeized gluon propagator combined with the tree-level
quark-reggeized gluon coupling:
\begin{align}
\label{eq:subtract_virtual}
& \mathcal{C}^{(1)}_{qr^* \to q}  \left(  p_a^+, \rho; \epsilon \frac{{\bm q}^2}{\mu^2} \right)  = \parbox{2cm}{\includegraphics[width = 2cm]{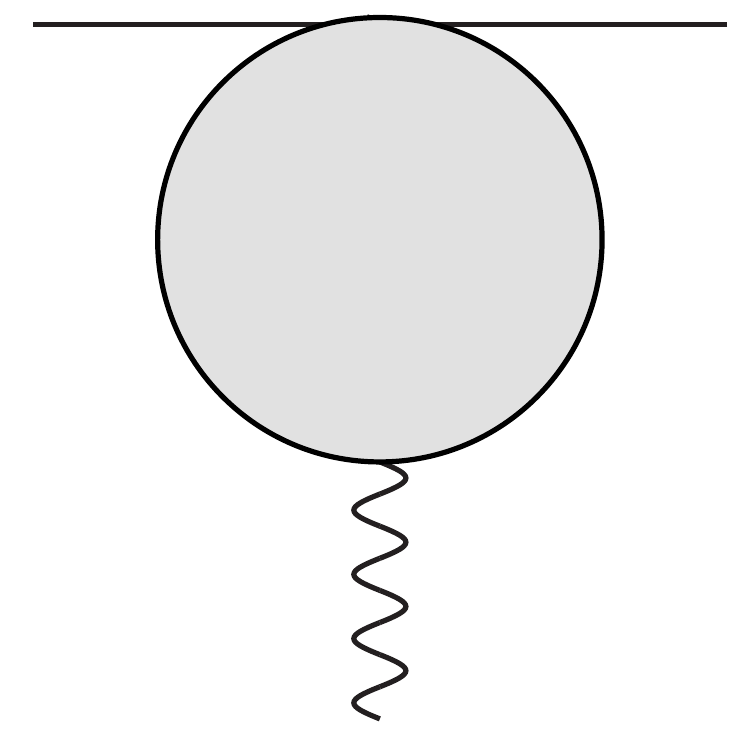}} =  
\parbox{2cm}{\includegraphics[width = 2cm]{impaamp.pdf}} 
- \parbox{2cm}{\includegraphics[width = 2cm]{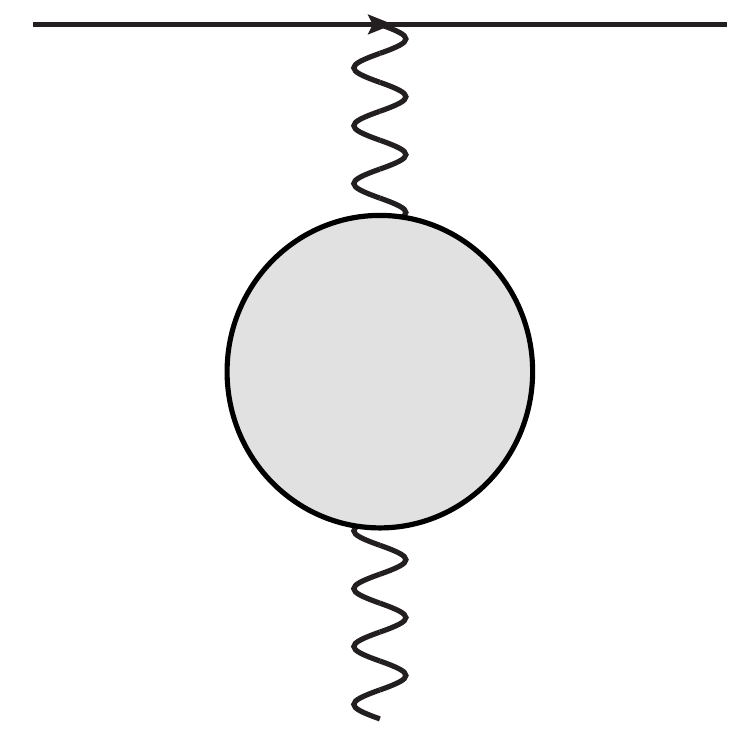}} \notag\\
&   = i  \mathcal{M}^{(0)}_{qr^* \to q}   \frac{\alpha_s  }{4\pi}  
\left(\frac{ {\bm k}^2}{\mu^2} \right)^\epsilon  \frac{ \Gamma^2 (1+\epsilon)}{\Gamma(1+2 \epsilon)}  \left\{
 \frac{-2 N_c}{\epsilon}  
\left(  \ln \frac{p_a^+}{\sqrt{{\bm k}^2}} - \frac{\rho}{2} \right)  \right. \notag \\
 & +  \left. 
\frac{ N_c (2 + 7 \epsilon)}{2\epsilon^2(1 + 2\epsilon)} + 
 \frac{1}{N_c} 
\left(  \frac{1}{\epsilon^2 (1 + 2 \epsilon)} + \frac{1}{2 \epsilon} \right)  - N_c\frac{1}{\epsilon}  \bigg(\psi(1-\epsilon) - 2 \psi(\epsilon) + \psi(1)\bigg)\right\}.
\end{align}
The four-point elastic amplitude is the sum of two contributions as the one calculated above:
\begin{align}   
i\mathcal{M}_{q_a q_b \to q_1 q_2}^{(1)} =
\parbox{2cm}{\includegraphics[width = 2cm]{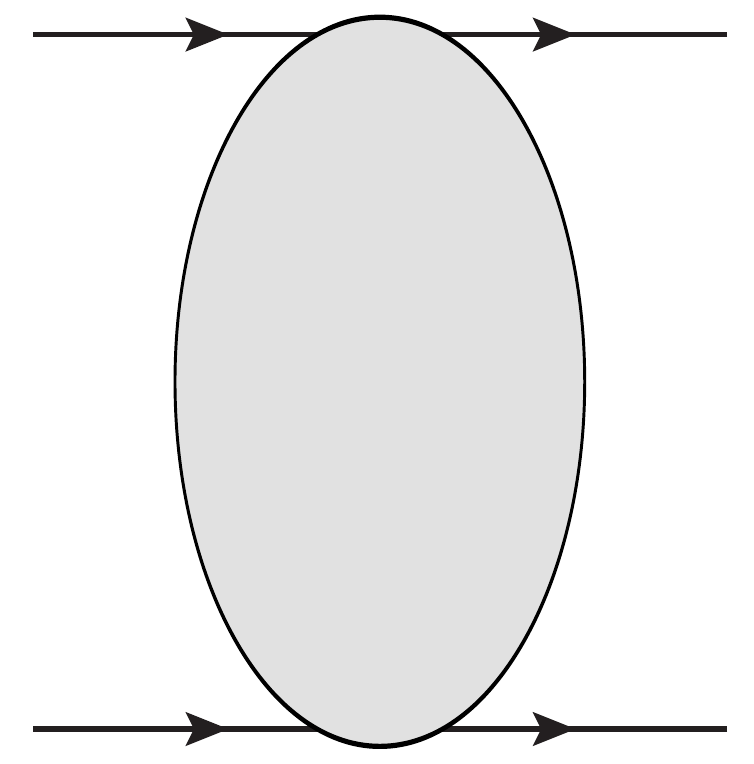}}
&= 
\parbox{2cm}{\includegraphics[width = 2cm]{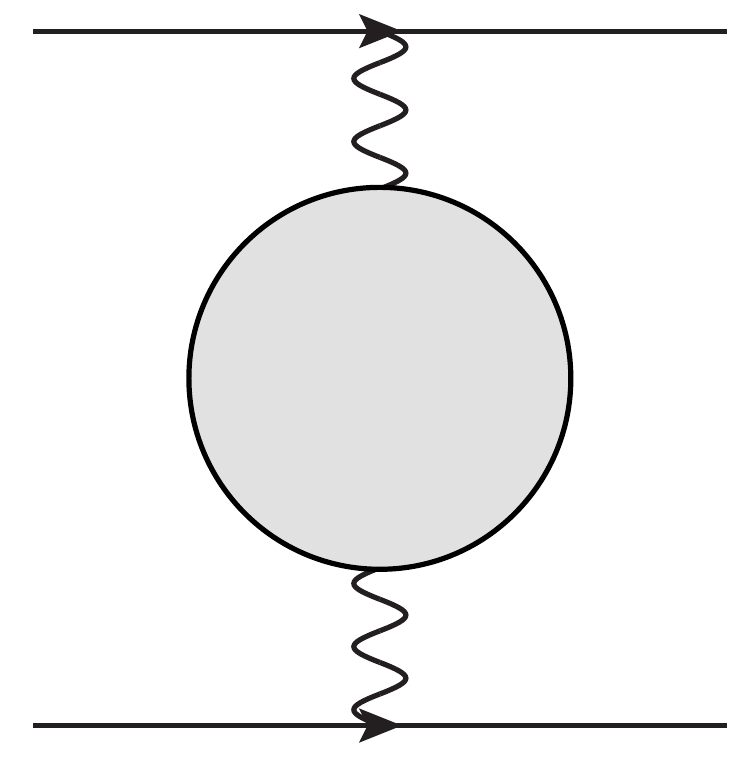}}
+
\parbox{2cm}{\includegraphics[width = 2cm]{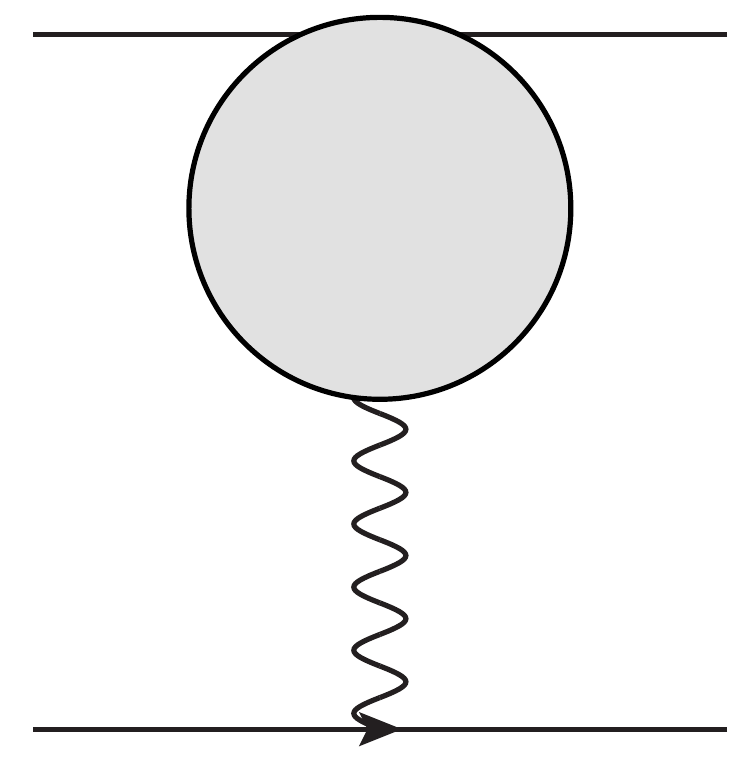}}
+
\parbox{2cm}{\includegraphics[width = 2cm]{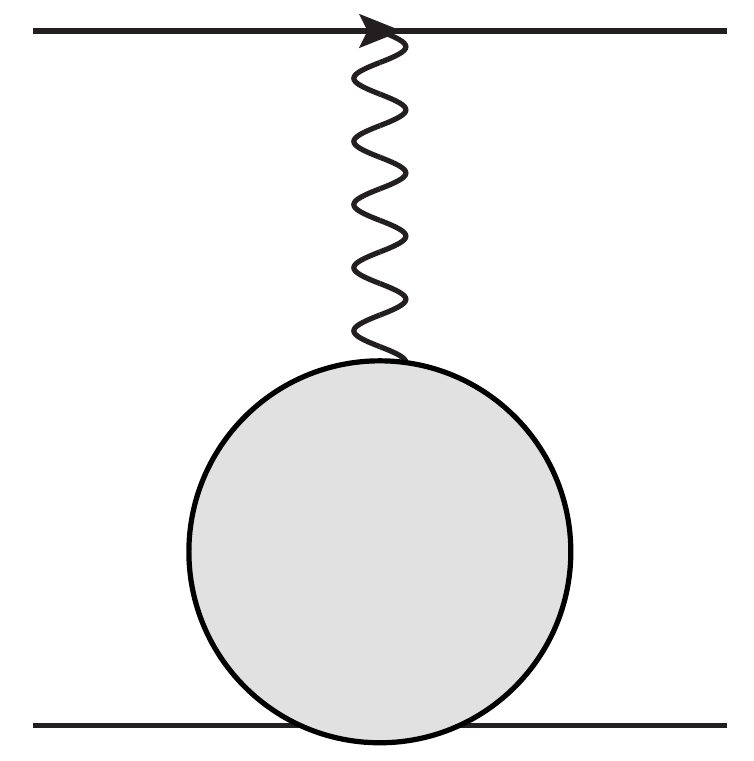}}. 
\label{eq:1loopgra}
\end{align}
While each of the three diagrams at the right hand side of Eq.~\eqref{eq:1loopgra} is divergent in the 
$\rho \to \infty$ limit,  these $\rho$-divergences cancel in the sum, yielding the following result:
\begin{align}\label{eq:1loopres}
\frac{i\mathcal{M}_{q_a q_b \to q_1 q_2}^{(1)}}{ i\mathcal{M}_{q_a q_b \to q_1 q_2}^{(0)}  } &= 
\frac{1}{2} \left( \ln \frac{s}{ {\bm k}^2}  +  \ln \frac{-s}{ {\bm k}^2 } \right) \omega^{(1)}\left(\epsilon, \frac{ {\bm q}^2}{\mu^2}\right)  
+ \Gamma_a^{(1)} \left(\epsilon, \frac{ {\bm q}^2}{\mu^2}\right)  + \Gamma_b^{(1)}\left(\epsilon, \frac{ {\bm q}^2}{\mu^2}\right) .
\end{align}
Here the one-loop gluon Regge 
trajectory reads
\begin{align}
  \label{eq:omega1}
   \omega^{(1)}\left(\epsilon, \frac{ {\bm q}^2}{\mu^2}\right) & =  -\frac{\alpha_s  N_c }{2\pi \epsilon}  
\left(\frac{ {\bm k}^2}{\mu^2} \right)^\epsilon  \frac{ \Gamma^2 (1+\epsilon)}{\Gamma(1+2 \epsilon)}.
\end{align}
 This piece is associated to the reggeized gluon exchange, whereas 
\begin{align}
\label{eq:G1}
 \Gamma^{(1)}_{a,b}\left(\epsilon, \frac{ {\bm q}^2}{\mu^2}\right)   &=
\frac{\alpha_s}{2} \left(\frac{{\bm k}^2}{\mu^2} \right)^\epsilon 
\bigg[\frac{N_c}{\pi}  \frac{85 + 9 \pi^2}{36} -\frac{\beta_0}{4 \pi \epsilon} - \frac{5}{18} \frac{n_f}{\pi} \notag \\
& \qquad \qquad  \qquad \qquad 
-\frac{C_F}{\pi} \left(  \frac{1}{\epsilon^2}  - \frac{3}{2\epsilon} + 4 -  \frac{\pi^2}{6}\right) \bigg] + \mathcal{O}(\epsilon),
\end{align}
provides the virtual corrections to the quark impact factor.  This result
is in perfect agreement with the QCD calculations
by~\cite{Fadin:1993qb}, confirmed in~\cite{DelDuca:1998kx}.  At this
point we would like to stress that in order to arrive at this result
it is necessary to subtract in Eq.~\eqref{eq:subtract_virtual} not
only the divergent pieces $\sim \rho$ but also finite terms.  This is
in contrast with the real contributions, where the entire central corrections
are proportional to the high energy divergence. We now discuss the calculation of the gluon Regge 
trajectory in terms of the quark contributions. 

\section{Quark contribution to the NLO gluon Regge trajectory}
\label{sec:renom}

It is useful to approach the calculation of the gluon Regge trajectory from the point of view of renormalized 
effective vertices and propagators. Formally, this can be understood as a renormalization of the 
coefficients of the reggeized gluon action as obtained after 
integrating out gluon and quark fields with the subsequent subtraction of non-local contributions. At the amplitude level this corresponds to the following definition of the renormalized quark-reggeized gluon coupling coefficients:
\begin{align}
  \label{eq:renomcoeff}
   \mathcal{C}^{ \text{R}}_{qr^* \to q} \left( \frac{ p_a^+}{M^+}; \epsilon,  \frac{{\bm q}^2}{\mu^2} \right) 
& =
 Z^+ \left( \frac{M^+}{\sqrt{{\bm k}^2}}, \rho ; \epsilon,  \frac{{\bm k}^2}{\mu^2}  \right)   \mathcal{C}_{qr^* \to q} \left(  \frac{ p_a^+}{\sqrt{{\bm k}^2}}, \rho; \epsilon \frac{{\bm k}^2}{\mu^2} \right), 
\\
   \mathcal{C}^{ \text{R}}_{qr^* \to q} \left( \frac{ p_b^-}{M^-}; \epsilon,  \frac{{\bm k}^2}{\mu^2} \right) 
& =
 Z^- \left( \frac{M^-}{\sqrt{{\bm k}^2}}, \rho ; \epsilon,  \frac{{\bm k}^2}{\mu^2}  \right)   \mathcal{C}_{qr^* \to q} \left(  \frac{ p_b^-}{\sqrt{{\bm k}^2}}, \rho; \epsilon \frac{{\bm k}^2}{\mu^2} \right), 
\end{align}
and the renormalized reggeized gluon propagator:
\begin{align}
  \label{eq:renompropR}
  G^{\text{R}} \left(M^+, M^-; \epsilon, {\bm k}^2, \mu^2 \right)  & = 
\frac{ G \left(\rho; \epsilon, {\bm k}^2, \mu^2   \right) }{ Z^+ \left( \frac{M^+}{\sqrt{{\bm k}^2}}, \rho ; \epsilon,  \frac{{\bm k}^2}{\mu^2}  \right) Z^- \left( \frac{M^-}{\sqrt{{\bm k}^2}}, \rho ; \epsilon,  \frac{{\bm k}^2}{\mu^2}  \right) },
\end{align}
with the bare reggeized gluon propagator given by
\begin{align}
  \label{eq:barepropR}
   G \left(\rho; \epsilon, {\bm k}^2, \mu^2   \right)
&=
\frac{i/2}{{\bm k}^2} \left\{ 1 + \frac{i/2}{{\bm k}^2}
 \Sigma \left(\rho; \epsilon, \frac{{\bm k}^2}{\mu^2}    \right)  + \left[  \frac{i/2}{{\bm k}^2} \Sigma \left(\rho; \epsilon, \frac{{\bm k}^2}{\mu^2}   \right)\right] ^2 + \ldots   \right\}. 
\end{align}
The renormalization factors $Z^\pm$ cancel for the complete quark-quark scattering amplitude and can be  parameterized as follows
\begin{align}
  \label{eq:Z+-}
  Z^\pm\left( \frac{M^\pm}{\sqrt{{\bm k}^2}}, \rho ; \epsilon,  \frac{{\bm k}^2}{\mu^2}  \right)  & = \exp \left [ \left(\rho -  \ln \frac{M^\pm}{\sqrt{\bm k}^2}  \right) \omega\left(\epsilon, \frac{{\bm k}^2}{\mu^2} \right) + f\left(\epsilon, \frac{{\bm k}^2}{\mu^2} \right) \right]
\end{align}
where the gluon Regge trajectory has the following perturbative expansion,
\begin{align}
  \label{eq:omega_expand}
   \omega\left(\epsilon, \frac{{\bm k}^2}{\mu^2} \right) & =  \omega^{(1)}\left(\epsilon, \frac{{\bm k}^2}{\mu^2} \right) + \omega^{(2)}\left(\epsilon, \frac{{\bm k}^2}{\mu^2} \right) + \ldots,
\end{align}
with the one-loop expression given in Eq.~\eqref{eq:omega1}.
The function $f(\epsilon, {\bm k}^2)$ parametrizes finite contributions and is, 
in principle, arbitrary.  Regge theory suggests to fix it in such a way that at one loop the non-$\rho$-enhanced contributions of the one-loop reggeized gluon self energy are entirely transferred to the
quark-reggeized gluon couplings leading to 
\begin{align}
  \label{eq:f1loop}
   f^{(1)}\left(\epsilon, \frac{{\bm k}^2}{\mu^2} \right) & =  \frac{ \alpha_s N_c \Gamma^2(1 + \epsilon)}{4 \pi \Gamma(1 + 2 \epsilon)} \left(\frac{{\bm q}^2}{\mu^2} \right)^\epsilon  
         \frac{(-1)}{(1 + 2 \epsilon)2 \epsilon} \bigg[   2 
 +  \frac{5 + 3\epsilon}{3 + 2 \epsilon} 
-\frac{n_f}{N_c} \left( \frac{2 + 2\epsilon}{3 + 2\epsilon} \right)\bigg]  .
\end{align}
Using $M^+ = p_a^+$, $M^- = p_b^-$ we can see that this choice for $f$ keeps the full $s$-dependence of 
the amplitude  inside the reggeized gluon exchange, while  the renormalized
quark-reggeized gluon couplings  agree with 
Eq.~\eqref{eq:G1}.
\begin{figure}
  \parbox{1.5cm}{\vspace{0.1cm} \includegraphics[height = 2.5cm]{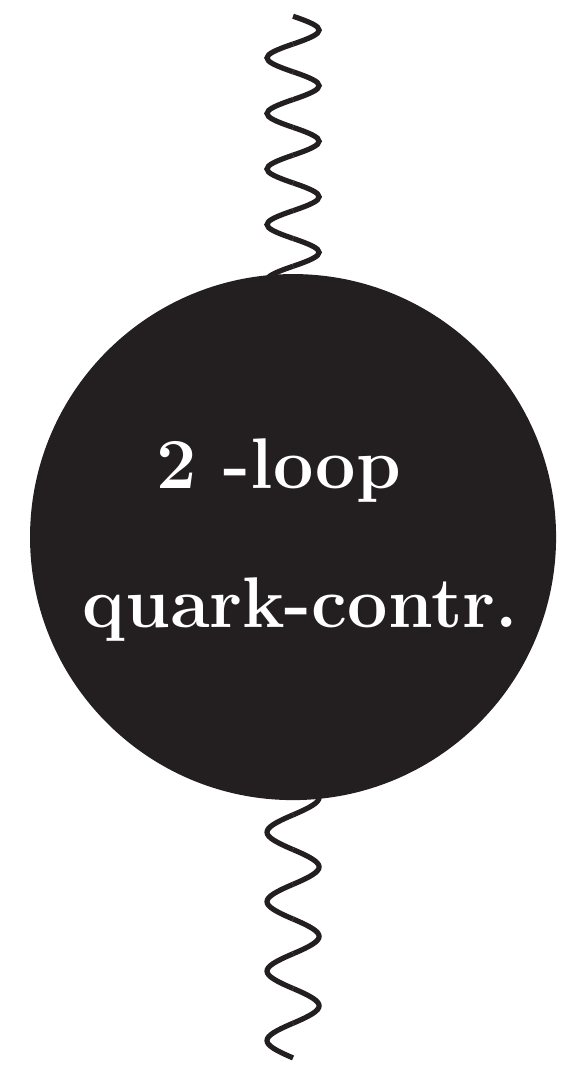}}
 = 
\parbox{1.4cm}{\vspace{0.1cm} \includegraphics[height = 2.5cm]{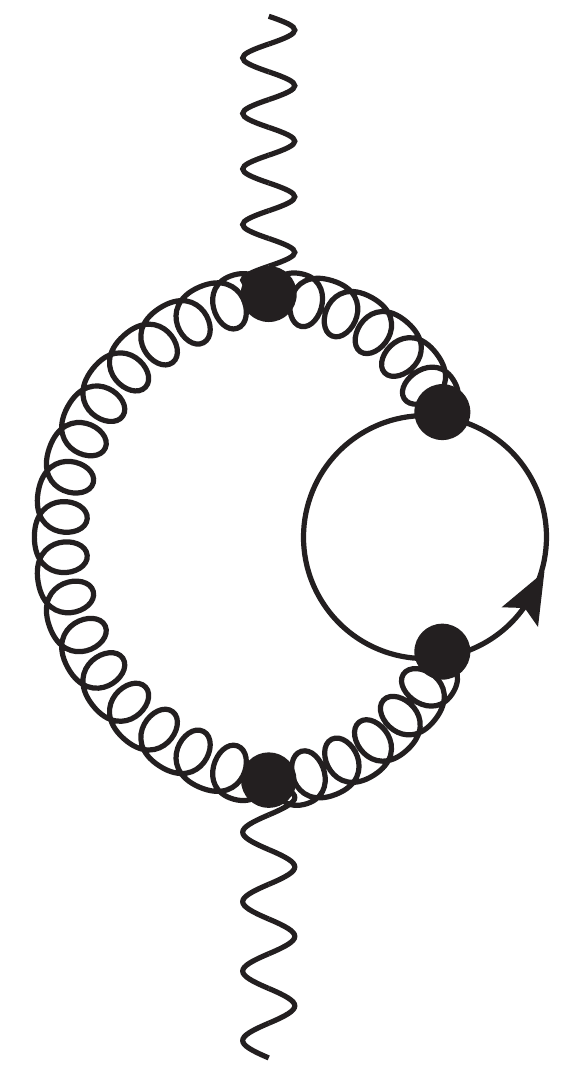}}
+
\parbox{1.4cm}{\vspace{0.1cm} \includegraphics[height = 2.5cm]{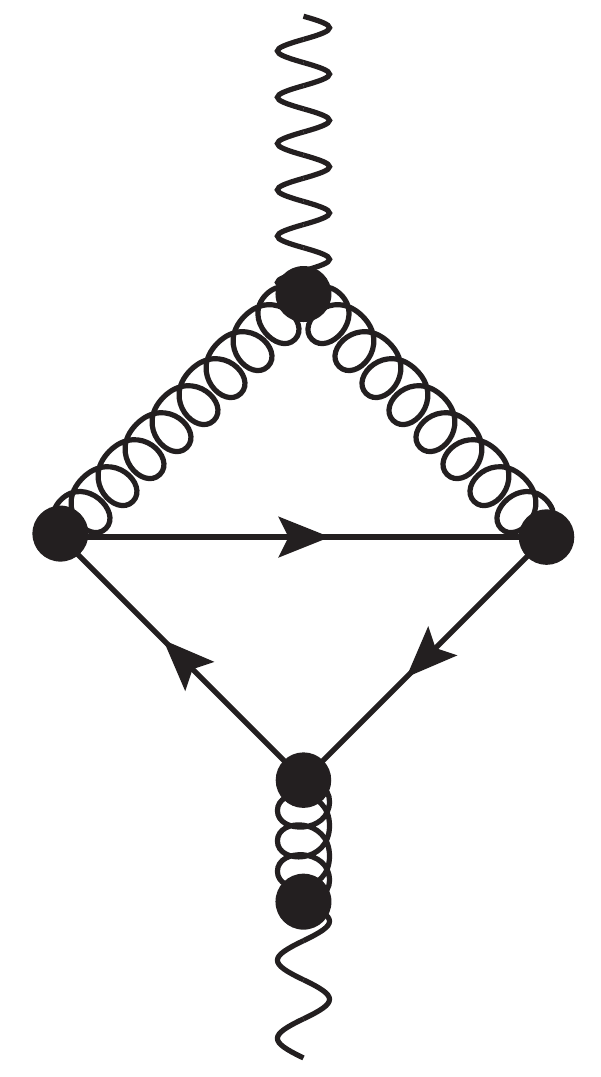}}
+
\parbox{1.4cm}{\vspace{0.1cm} \includegraphics[height = 2.5cm]{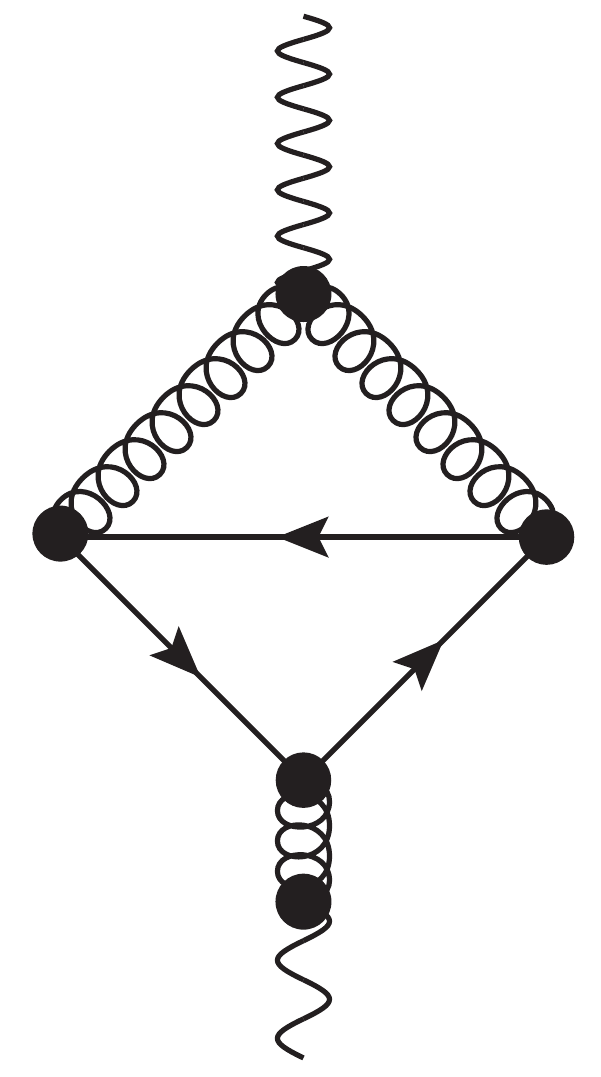}}
+
\parbox{1.4cm}{\vspace{0.1cm} \includegraphics[height = 2.5cm]{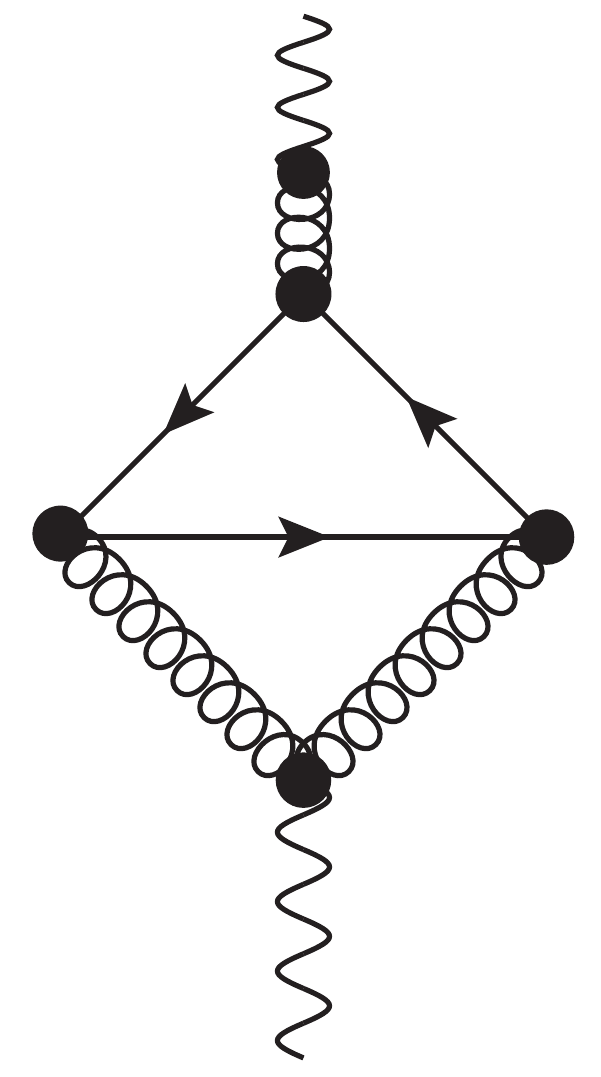}}
+
\parbox{1.4cm}{\vspace{0.1cm} \includegraphics[height = 2.5cm]{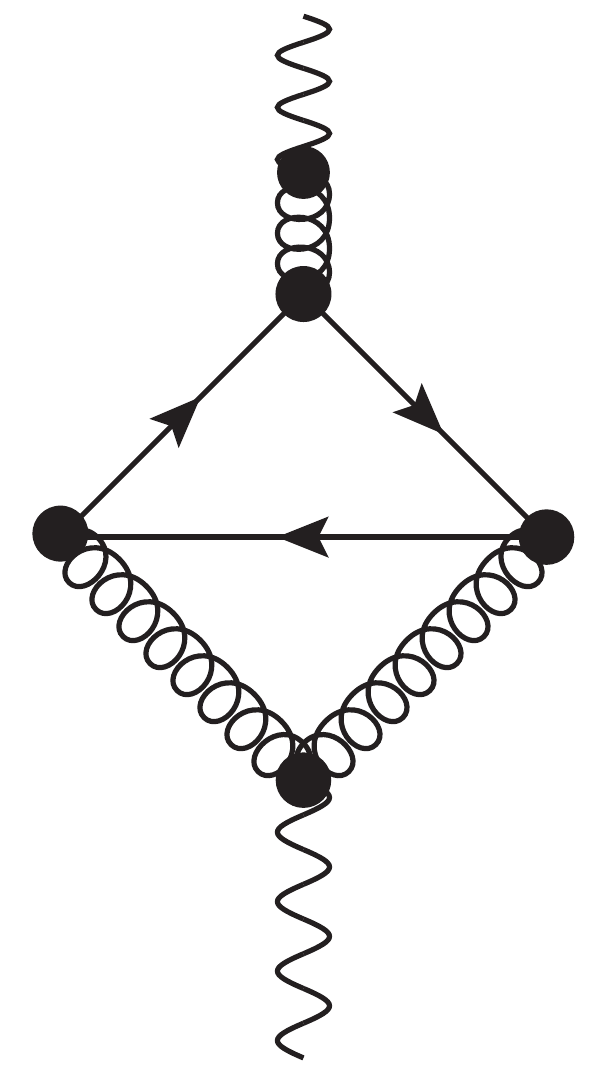}}
+
\parbox{1.4cm}{\vspace{0.1cm} \includegraphics[height = 2.5cm]{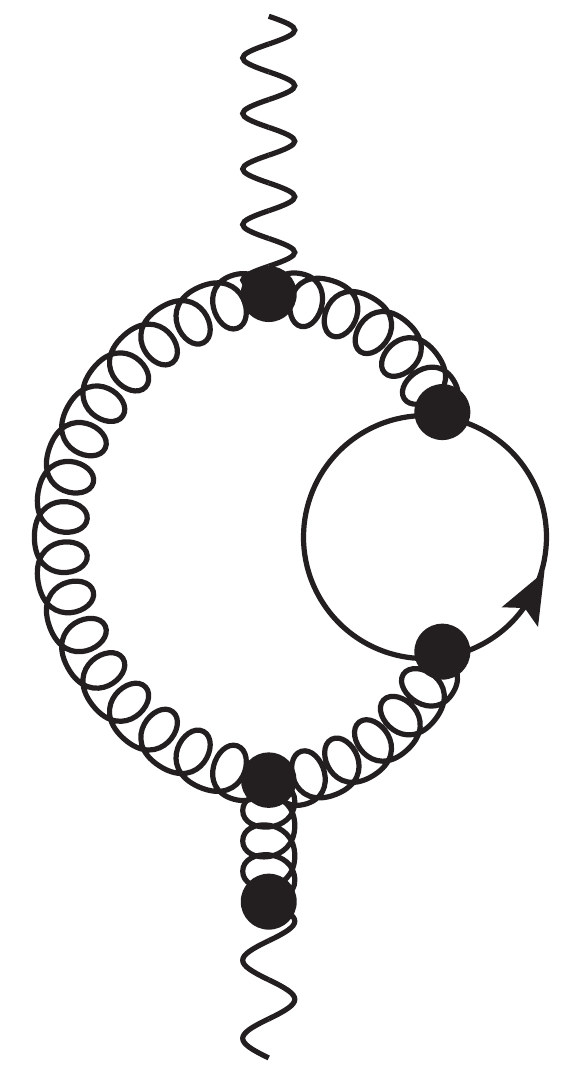}}
+
\parbox{1.4cm}{\vspace{0.1cm} \includegraphics[height = 2.5cm]{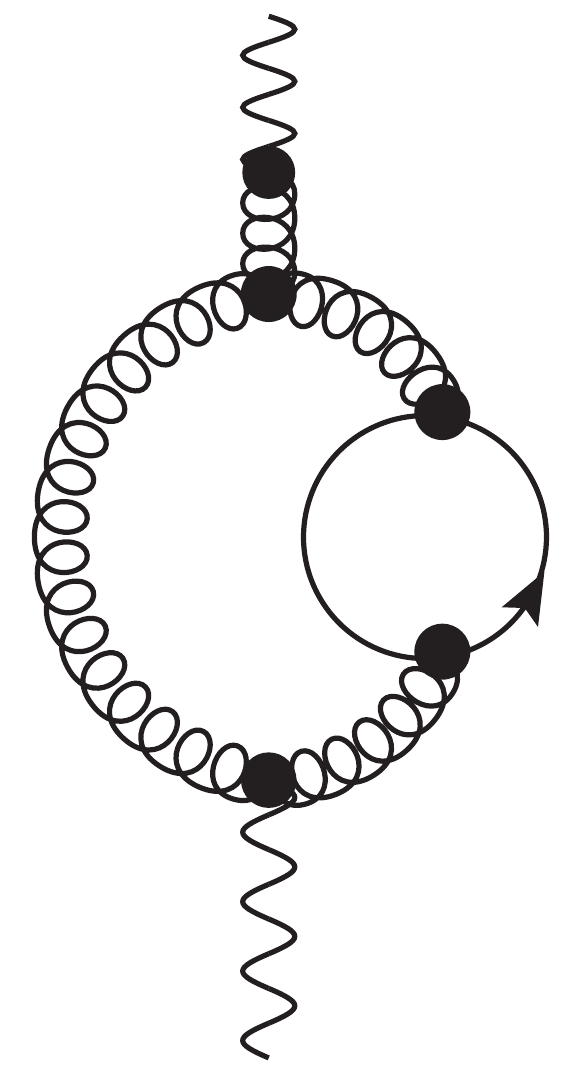}}
+
\parbox{1.4cm}{\vspace{0.1cm} \includegraphics[height = 2.5cm]{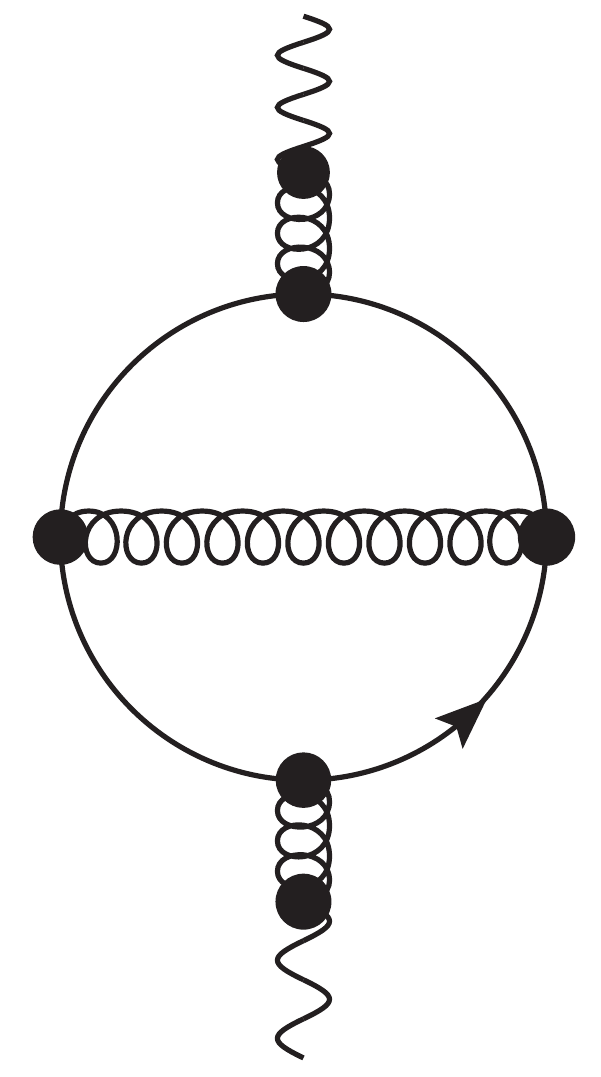}}
+
\parbox{1.4cm}{\vspace{0.1cm} \includegraphics[height = 2.5cm]{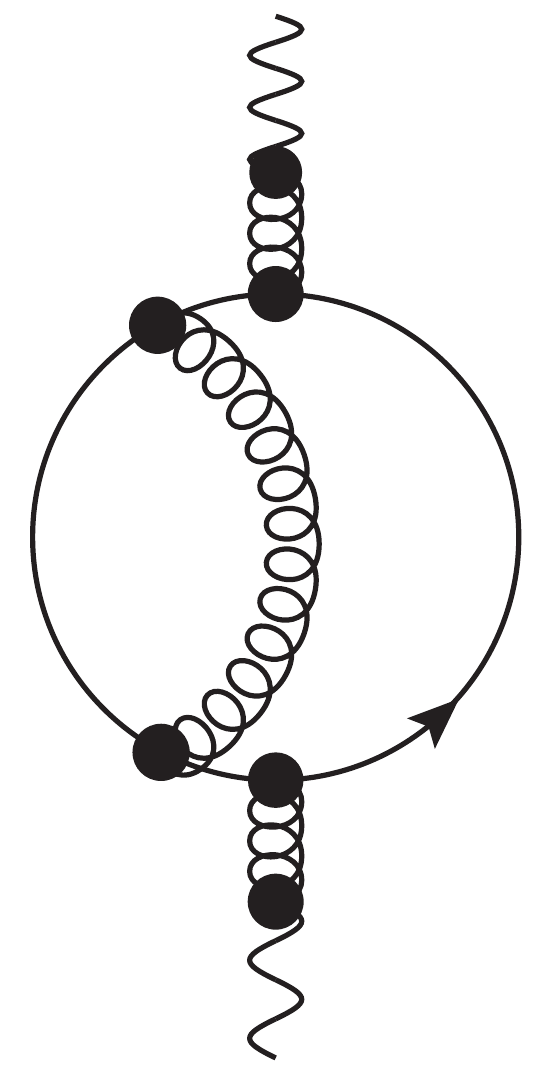}}
+
\parbox{1.4cm}{\vspace{0.1cm} \includegraphics[height = 2.5cm]{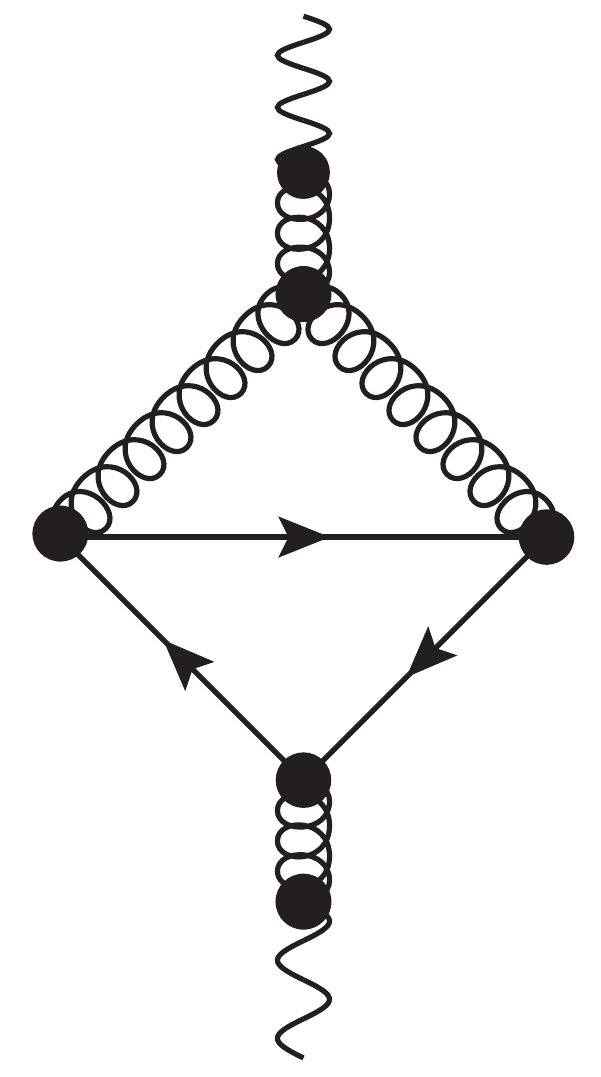}}
+
\parbox{1.4cm}{\vspace{0.1cm} \includegraphics[height = 2.5cm]{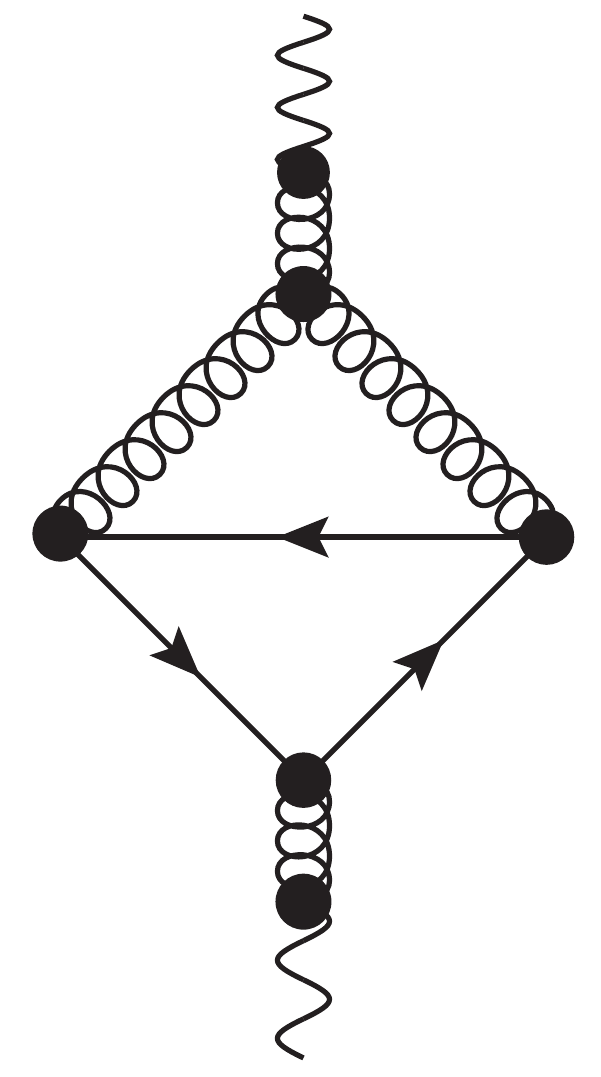}}
+
\parbox{1.4cm}{\vspace{0.1cm} \includegraphics[height = 2.5cm]{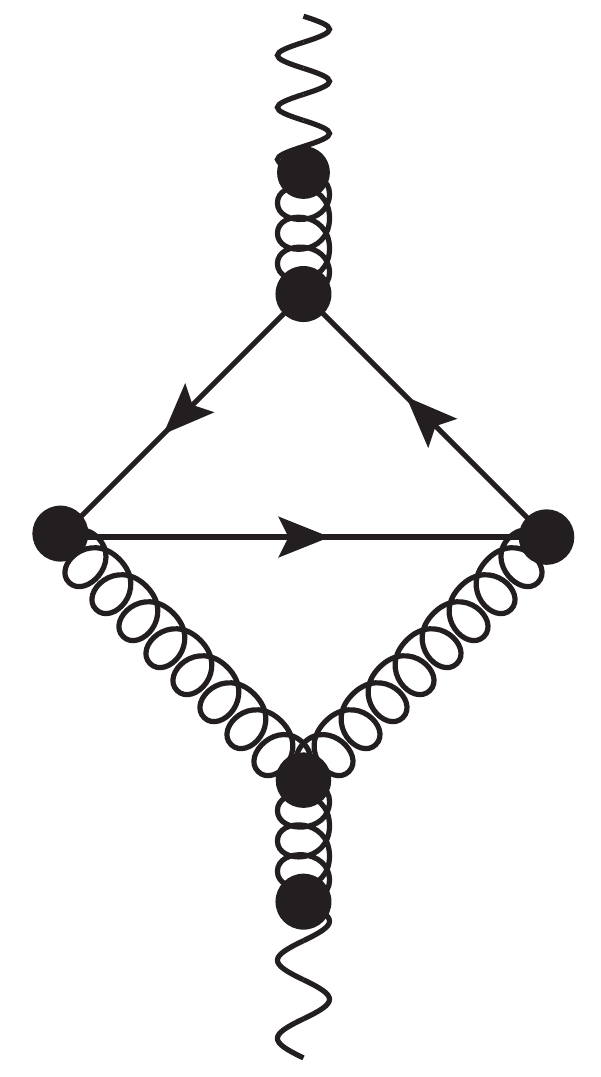}}
+
\parbox{1.4cm}{\vspace{0.1cm} \includegraphics[height = 2.5cm]{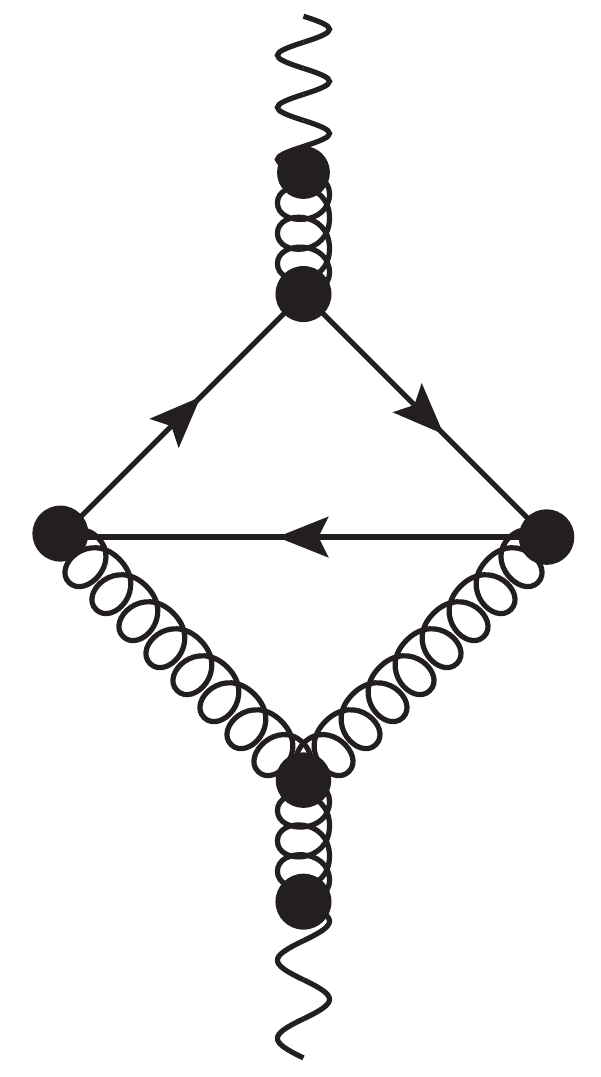}}
+
\parbox{1.4cm}{\vspace{0.1cm} \includegraphics[height = 2.5cm]{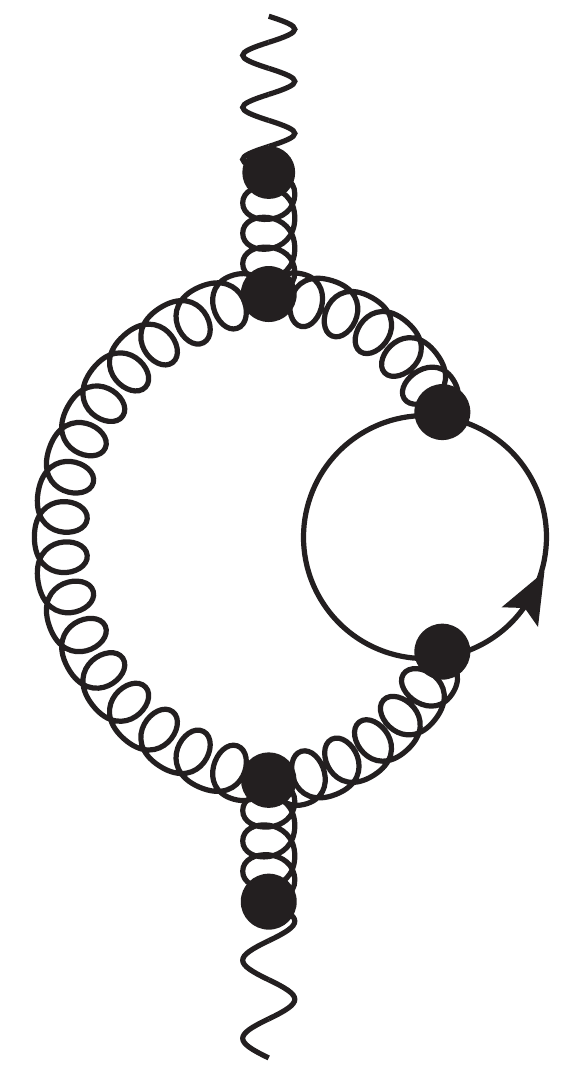}}
\caption{\small Diagrams contributing to the one-loop reggeized gluon self-energy.}
\label{fig:self_2loop}
\end{figure}
In \cite{Chachamis:2012gh} this formalism has been put to test through the determination of the quark contributions to the two-loop gluon Regge trajectory. 
The complete set of contributing diagrams is given in
Fig.~\ref{fig:self_2loop}. The diagrams in the second line of this figure can be neglected since they 
do not give $\rho$-enhanced contributions. These subleading diagrams are obtained as projections of
the quark contribution to the two-loop gluon polarization tensor and are
therefore finite when $\rho \to \infty$.  

Among all the remaining diagrams we found that only the first one in Fig.~\ref{fig:self_2loop} is 
$\rho$-enhanced. This graph is unique since the reggeized gluon couples from above and below  
to the usual gluon loop through an induced vertex. 
The complete set of enhanced contributions in Fig.~\ref{fig:self_2loop} is given by
\begin{align}
 \parbox{1.5cm}{\vspace{0.1cm} \includegraphics[height = 2.5cm]{qself2L.pdf}}   & =
-\rho  (-i 2 {\bm k}^2) \bar{g}^4 \frac{4 n_f}{\epsilon N_c}  \frac{\Gamma^2(2 + \epsilon)}{\Gamma(4 + 2\epsilon)} \cdot \frac{3 \Gamma(1 - 2\epsilon) \Gamma(1 + \epsilon) \Gamma(1 + 2\epsilon)}{\Gamma^2(1 - \epsilon) \Gamma(1 + 3 \epsilon) \epsilon}.
  \label{eq:traj_jose}
\end{align}
As for the one-loop corrections to the quark-quark-reggeized gluon vertex, it is needed to subtract from this result the corresponding diagrams with a reggeized gluon exchange. For the complete two-loop trajectory we have
\begin{align}
  \label{eq:coeff_2loop}
  \parbox{2cm}{\center \includegraphics[height = 2.5cm]{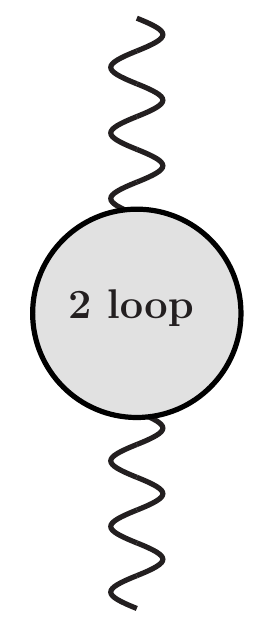}}  
  = 
  \parbox{2cm}{\center \includegraphics[height = 2.5cm]{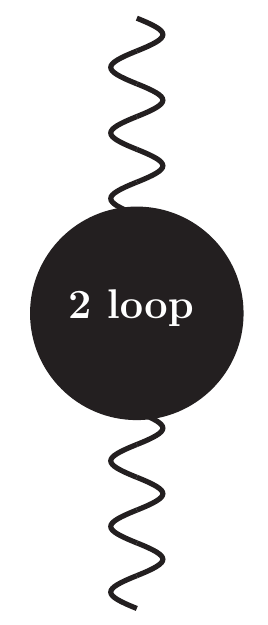}}
  -
  \parbox{2cm}{\center \includegraphics[height = 2.5cm]{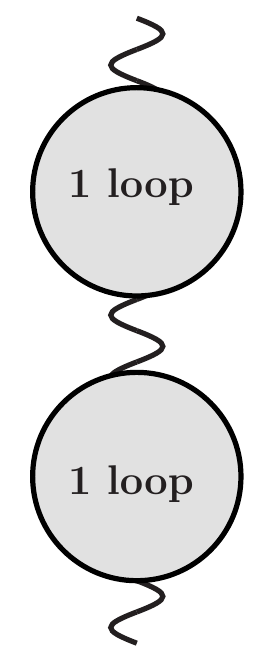}}.
\end{align}
More precisely, the terms to be subtracted which are both proportional to $n_f$ and $\rho$-enhanced read
\begin{align}
  \label{eq:doubleloo}
  \left[ \parbox{1.5cm}{\center \includegraphics[height = 2.5cm]{self2_1loop.pdf}}
\right]_{n_f, \,\rho}  
&= \hspace{0.4cm}
- \rho( -2i{\bm k}^2) \frac{\alpha_s^2 N_c}{(4 \pi)^2}  \left( \frac{{\bm k}^2}{\mu^2} \right)^{2 \epsilon} \frac{8 n_f}{\epsilon} \frac{ \Gamma^2(2 + \epsilon) }{ \Gamma(4 + 2 \epsilon)}\frac{\Gamma^2(1 + \epsilon)}{ \Gamma(1 + 2\epsilon)}\frac{2}{\epsilon}.
\end{align}
The final subtracted reggeized gluon self energy, in terms of $n_f$ and $\rho$ contributions is
\begin{align}
 \Sigma^{(2)}_{n_f} \left(\rho; \epsilon, \frac{{\bm k}^2}{\mu^2}    \right) & = 
 \left[   \parbox{1.5cm}{\center \includegraphics[height = 2.5cm]{self_2loop.pdf}} \right]_{n_f, \rho}  
 =
    \frac{\rho( -2i{\bm k}^2) \alpha_s^2 N_c 4 n_f}{ (4 \pi)^2   \epsilon} \frac{ \Gamma^2(2 + \epsilon) }{ \Gamma(4 + 2 \epsilon)}
 \left( \frac{{\bm k}^2}{\mu^2} \right)^{2 \epsilon} 
\notag \\
& \qquad \qquad \qquad \times 
 \left(\frac{\Gamma^2(1 + \epsilon)}{ \Gamma(1 + 2\epsilon)}\frac{4}{\epsilon}  -\frac{3 \Gamma(1 - 2\epsilon) \Gamma(1 + \epsilon) \Gamma(1 + 2\epsilon)}{\Gamma^2(1 - \epsilon) \Gamma(1 + 3 \epsilon) \epsilon}\right).
     \label{eq:traje_quark_coeff}
\end{align}
To extract the quark contribution to the gluon Regge trajectory at two loops it is needed to identify 
the $\rho$-enhanced contributions of the bare reggeized gluon propagator up to two loops. These are 
\begin{align}
  \label{eq:GBare2loop}
   G \left(\rho; \epsilon, {\bm k}^2, \mu^2   \right) & = \frac{i/2}{{\bm k}^2} \bigg\{ 1 + \frac{i/2}{{\bm k}^2} \Sigma^{(1)} \left(\rho; \epsilon, \frac{{\bm k}^2}{\mu^2}    \right) 
\notag \\
&  
+ 
\frac{i/2}{{\bm k}^2} \Sigma^{(2)} \left(\rho; \epsilon, \frac{{\bm k}^2}{\mu^2}    \right)
+
\left[  \frac{i/2}{{\bm k}^2} \Sigma^{(1)} \left(\rho; \epsilon, \frac{{\bm k}^2}{\mu^2}  \right)   \right]^2 + \ldots   \bigg\}. 
\end{align}
Selecting only the terms proportional to $n_f$, the pieces of ${\cal O} (\rho)$ in the renormalized 
reggeized gluon propagator at two loops are
\begin{align}
  \label{eq:GRrho}
   G^{\text{R}} \bigg|_{n_f, \rho}
 & =  \frac{i/2}{{\bm k}^2} \bigg\{  \frac{i/2}{{\bm k}^2} \Sigma^{(2)}_{n_f} +  
\left[ 
 \frac{i/2}{{\bm k}^2} 
\Sigma^{(1)} \left(\rho;\epsilon, \frac{{\bm k}^2}{\mu^2} \right)   \right]^2 \bigg|_{n_f, \rho}
\notag \\
&  \qquad \qquad 
- \rho \,  \omega^{(1)}\left(\epsilon, \frac{{\bm k}^2}{\mu^2} \right) f^{(1)}_{n_f} \left(\epsilon, \frac{{\bm k}^2}{\mu^2} \right) -  \rho \, \omega^{(2)}_{n_f}\left(\epsilon, \frac{{\bm k}^2}{\mu^2} \right) \bigg\}
\end{align}
with
\begin{align}
  \label{eq:fnf}
  f^{(1)}_{n_f} \left(\epsilon, \frac{{\bm k}^2}{\mu^2} \right) & =   n_f \frac{ \alpha_s   \Gamma^2(1 + \epsilon)}{ 4 \pi \Gamma(1 + 2 \epsilon)} \left(\frac{{\bm q}^2}{\mu^2} \right)^\epsilon  
  \bigg(\frac{2 + 2\epsilon}{3 + 2\epsilon}\bigg) 
\end{align}
being the part proportional to $n_f$ of $f^{(1)}$ in Eq.~\eqref{eq:f1loop}. 
The requirement that the $\rho$ dependence in Eq.~\eqref{eq:GRrho} has to cancel in the limit 
$\rho \to \infty$ then yields the quark contribution to the gluon Regge trajectory at two loops as
\begin{align}
  \label{eq:traje_quark_coeff}
 \omega^{({2})}_{n_f}\left(\epsilon, \frac{{\bm k}^2}{\mu^2} \right)  & = 
 \frac{\alpha_s^2 N_c}{4 \pi^2}  
\left( \frac{{\bm k}^2}{\mu^2} \right)^{2 \epsilon} \frac{ n_f}{\epsilon} \frac{ \Gamma^2(2 + \epsilon) }{ \Gamma(4 + 2 \epsilon)}
\notag \\
& 
\qquad   \times  \bigg [ \frac{\Gamma^2(1 + \epsilon)}{ \Gamma(1 + 2\epsilon)}\frac{2}{\epsilon} -\frac{3 \Gamma(1 - 2\epsilon) \Gamma(1 + \epsilon) \Gamma(1 + 2\epsilon)}{\Gamma^2(1 - \epsilon) \Gamma(1 + 3 \epsilon) \epsilon}\bigg],
\end{align}
which is in perfect agreement with Eq.(9) of the original study of Fadin et al. in~\cite{Fadin:1996tb}. 
This confirms the validity of our approach 
to the effective action at two loop level and the regularization prescription here presented.

\section{Conclusions and Outlook}

In these contribution a brief introduction to Lipatov's effective
action to describe high energy processes in QCD has been given. This
effective action is not obtained through a reduction of the number of
degrees of freedom, but rather adds a new degree of freedom through
the reggeized gluon.  This leads to a series of subtleties that have
been recently addressed
in~\cite{Hentschinski:2011tz,Chachamis:2012gh,Chachamis:prep} by means
of a regularization and subtraction procedure that ensures the
locality in rapidity. Several non-trivial checks of the validity of
this prescription, including one- and two-loop computations, have been
performed finding agreement with previous results in the
literature. We have explicitly discussed the quark-initiated forward
jet vertex and quark contributions to the gluon Regge trajectories at
next-to-leading order.  The gluon-initiated jet vertex and the gluon
contributions to the gluon trajectory will be presented elsewhere~\cite{Chachamis:prep}.

\section*{Acknowledgments} 

We acknowledge partial support from the European Comission under contract LHCPhenoNet (PITN-GA-2010-264564), the Comunidad de Madrid through Proyecto HEPHACOS ESP-1473, and MICINN (FPA2010-17747).  M.H. also acknowledges support from the German Academic Exchange Service
(DAAD), the U.S. Department of Energy under contract number DE-AC02-98CH10886 and a BNL ``Laboratory Directed Research and Development'' grant (LDRD 12-034).


\end{document}